\documentclass{aa}
\usepackage{threeparttable}
\usepackage{longtable}
\usepackage{threeparttablex}
\usepackage{booktabs} 
\usepackage{tablefootnote}
\usepackage{lscape} 
\usepackage{epstopdf}
\usepackage{xcolor}
\usepackage{xspace}
\usepackage{graphicx}
\usepackage{amssymb}
\usepackage{amsmath}
\usepackage{natbib}
\usepackage{float}
\usepackage{hyperref}
\usepackage{soul}
\usepackage{verbatim}  % in the preamble
\newcommand{\TC}[1]{{\color{orange} #1}}

\begin{document}

    \title{What Drives the Bimodal Distribution of Eddington-Scaled Radio Luminosity in Nearby Early-Type Galaxies?}

   \subtitle{}
   \titlerunning{Bimodality in the Radio Luminosity of Early-Type Galaxies}
   \authorrunning{Wójtowicz et al.}

    \author{A. W\'ojtowicz
        \inst{1}
        \and
        N. Werner
        \inst{1}
        \and
        Ł. Stawarz
        \inst{2}        
        \and
        C.C. Cheung
        \inst{3}
      }
                
\institute{Department of Theoretical Physics and Astrophysics, Faculty of Science, Masaryk University, Kotláršká 2, Brno, 61137, Czechia\\
           \email{awojtowicz@oa.uj.edu.pl}
        \and           
           Astronomical Observatory, Jagiellonian University, ul. Orla 171, PL-30244 Krak\'ow, Poland\\
        \and
           Space Science Division, Naval Research Laboratory, Washington, DC 20375, USA
           }
           
   \date{Received ; accepted }

% \abstract{}{}{}{}{} 
% 5 {} token are mandatory
 
  \abstract
    { Early-type galaxies host low-luminosity active galactic nuclei, indicated by radio emission that can span several orders of magnitude in physical scale, ranging from compact parsec-scale to extended kpc-scale radio morphology. }
   {We investigate the Eddington-scaled radio luminosity distribution of nearby early-type galaxies in a large sample of 117 sources to confirm if this distribution is bimodal, as inferred based on a sample of 62 galaxies in \cite{Wojtowicz2023}, and whether this bimodality can be attributed to the specific host galaxy properties.}
   {We compile a sample of early-type galaxies with black hole masses measured using direct methods, as well as the radio flux densities at 1.4\,GHz and 3\,GHz. We use statistical tests to confirm whether the Eddington-scaled radio luminosities are bimodal. We investigate the properties of radio-dim and -bright sources, assess the presence of extended jets with VLASS imaging, and examine host galaxy kinematics and central stellar structure.}
   {We confirm, using twice larger sample of all known 1.4\,GHz-detected early-type galaxies with directly measured black hole masses (117 galaxies), that the distribution of $L_{\rm 1.4\,GHz}/L_{\mathrm{Edd}}$ is bimodal, with characteristic antimode at $L_{\rm 1.4\,GHz}/L_{\mathrm{Edd}}\approx -8.6$, which disappears when considering the subset of sources with black hole masses obtained using the $M_{\rm BH}-\sigma_\star$ relation. The radio-bright peak is dominated by galaxies hosting resolved jets, while radio-dim systems show compact nuclear emission with many showing excess radio emission relative to that expected from star formation as indicated by the well-known FIR/radio correlation. Moreover, we find that radio-bright galaxies are primarily slow rotators with depleted stellar cores, whereas radio-dim galaxies are predominantly fast rotators.}
   {We show that nearby early-type galaxies with direct black hole mass measurements exhibit a clear bimodality in Eddington-scaled radio luminosity, separating radio-dim nuclei with compact emission from radio-bright systems hosting extended jets. The dichotomy correlates strongly with host-galaxy kinematics and central structure, suggesting that the ability to sustain jet production is governed primarily by galaxy assembly history and feeding mode rather than by black hole mass or accretion rate alone. In this term the radio output of radio-dim sources likely reflects modest, intermittent supplies of magnetized gas delivered to the accretion flow through stochastic processes, most plausibly the tidal disruption of giant-branch stars passing through the immediate vicinity of the SMBH.
}

   \maketitle

\section{Introduction}
\label{sec:intro}

Elliptical galaxies can be classified into two main groups based on their observed properties. One group consists of massive ellipticals, typically with stellar masses exceeding $\sim 10^{11} M_{\odot}$. These galaxies often possess halos of hot, X-ray--emitting gas and are frequently associated with strong radio sources \citep{Bender1989}. Their luminosity profiles commonly exhibit core flattening and boxy isophotes \citep{Faber1997}. Kinematic studies show that massive ellipticals usually rotate slowly, often contain kinetically decoupled components, and display a notable amount of rotation along the galaxy minor axis \citep{Cappellari2007, Emsellem2007}. Less massive ellipticals, on the other hand, rarely exhibit detectable X-ray halos and typically show weak or no radio emission. Their luminosity profiles generally follow a smooth, power-law-like distribution, and their isophotes tend to be more disky. Kinematically, these less massive galaxies often show significant rotation along the galaxy’s major axis.

Elliptical galaxies are typical hosts of quasars, i.e., active galactic nuclei (AGN) accreting near the Eddington limit \citep[e.g.,][]{Dunlop2003, McLure1999, Floyd2004}. While all quasars, owing to the universality of accretion-disk emission, exhibit similar optical properties, their radio properties can vary substantially, with some quasars detected in the radio band and others not. A fundamental dichotomy in the underlying radio-emission mechanisms of quasars was first tentatively suggested in early studies by \citet{Kellermann1989}, who proposed two distinct classes --- `radio-quiet' and `radio-loud' quasars --- based on the bimodal distribution of the ratio of radio to optical flux densities.

%Since then, various takes on the radio dichotomy of quasar sources have been proposed in the literature, ranging from support for a bimodality \citep[e.g.,][]{Ivezic2002} to analysis favoring rather continuous distribution instead \citep[e.g.,][and references therein]{Singal2013}.

Complementing this dichotomy among quasar sources and extending it to the broader population of elliptical galaxies exhibiting varying levels of nuclear activity, large-sky radio surveys show that radio-emitting galaxies fall into two broad morphological classes at the resolution typical of such surveys: (a) resolved systems with extended jets and (b) unresolved, compact radio sources.

%The launching of the powerful outflow is well explained in a framework of the Blandford and Znajek mechanism \citep{BZ1977}, where the rotational energy of the black hole is extracted in the presence of a magnetic field embedded in the surrounding matter. In this model the power of the outflow is given as a function of matter supply (accretion rate), BH spin, and the strength of the magnetic field lines.

The origin of radio emission in extended radio galaxies is relatively well established: their powerful outflows are thought to be launched via the Blandford–Znajek mechanism, in which rotational energy is extracted from a spinning black hole threaded by magnetic fields anchored in the surrounding accreting matter \citep{BZ1977}.In this model, the outflow power is set by the mass supply (accretion rate), which regulates the magnetic flux threading the black hole horizon, and by the black hole spin. However, the physical mechanisms responsible for collimating and accelerating these outflows to their terminal bulk velocities remain a matter of debate.

The characteristics of radio emissions from compact radio galaxies continue to be somewhat enigmatic and shrouded in mystery. Radio emission, in principle, is mostly associated with synchrotron processes, where relativistic particles are accelerated in a turbulent magnetic field, preferentially at the fronts of strong shock waves. However, such conditions occur not only in AGN-driven winds or jets but also in star-forming regions, particularly in the central parts of galaxies, implying that both environments may contribute to the observed radio emission.

A comprehensive summary of the problem associated with radio emission in various types of AGN was presented in, e.g., the \citet{Blandford2019} and \citet{Tadhunter2016} reviews. 

Observation of X-ray binaries supports the importance of spin in launching the jet \citep{Narayan2012}, yet on the other hand, many active galaxies with measured high spin values are radio quiet \citep[see][for a recent review]{Reynolds2021}. Thus, it seems that high spin alone is not sufficient to launch powerful and collimated outflows, highlighting the importance of accretion rate and magnetic field accumulation. In most recent studies, particularly those based on numerical simulations \citep[see][for a review]{Komissarov2021}, jets are indeed found to form most efficiently in the ``magnetically arrested disk'' (MAD) regime. In this state, the poloidal magnetic flux accumulated near the black hole reaches saturation, enabling efficient extraction of rotational energy from the ergosphere and creating a rigid, magnetically dominated funnel around the jet base that facilitates its initial collimation. 

%One possible clue on the key mechanism in action can follow from the absence of powerful jets associated with spiral hosts. This seems to indicate that it is the environment beyond the Black Hole sphere of influence that plays the most significant role in magnetic field buildup. The large scale inflow in spiral galaxies is mostly equatorial, while in elliptical it is mostly quasi-spherical as it originates in the circumgalactic medium.
A comprehensive review of the radio emission mechanisms in radio-quiet AGN is provided by \citet{Panessa2019}.  In this context, building on our previous analysis presented in \citet{Wojtowicz2023}, we investigate the origin of the radio emission in a sample of early-type galaxies with accurately measured supermassive black-hole (SMBH) masses. These systems accrete at low or extremely low rates and are typically observed to be radio-faint.

In \cite{Wojtowicz2023}, based on a sample of 62 early-type galaxies with directly estimated BH masses, $M_{\rm BH}$, we identified a possible bimodality in a distribution of the logarithm of the integrated 1.4\,GHz radio luminosities expressed as a fraction of their Eddington luminosities, $\log L_{\mathrm{1.4\,GHz}}/L_{\mathrm{Edd}}$, where $\log L_{\mathrm{Edd}}/{\rm erg\,s^{-1}} \simeq 38.1+\log M_{\rm BH}/M_{\odot}$.

%That sample composing of 62 sources was insufficient to claim the bi-modality adopting the standard approach, such as \citep{aa2019} excess mass test. 

In this paper, we present a large sample of 123 early-type galaxies with significant radio detections at 1.4\,GHz. Six sources were subsequently excluded because they only have upper limits to their BH mass estimates, thus giving a net sample of 117 galaxies in our analysis. We test whether the bimodality identified in \citet{Wojtowicz2023} remains present in the expanded sample.

\section{Sample selection and Data Acquisition}
\label{sec:sample}
An extensive search of the up-to-date literature allowed us to compile a list of all galaxies with central SMBH masses measured using direct methods -- these include measurements through direct observations of gas and stellar kinematics, as well as from reverberation mapping.

Here we only focus and present the sample of early-type galaxies (i.e., lenticular or elliptical). These were classified through visual investigation of the available optical images in {\textit{Digital Sky Survey (DSS)}\footnote{\url{https://archive.eso.org/dss/dss}}}, {\textit{Sloan Digital Sky Survey (SDSS)\footnote{\url{https://www.sdss.org/dr18/}}}}, {\textit{The Panoramic Survey Telescope \& Rapid Response System (Pan-STARRS)\footnote{\url{https://catalogues.mast.stsci.edu/panstarrs/}}}}. In case of unclear morphology, we further investigate high-resolution images if available through {\it{Hubble Legacy Archive \footnote{\url{https://hla.stsci.edu/}}}}. A galaxy was classified as early-type if its luminosity distribution decreased smoothly from the center and showed no signs of spiral arms, or disturbance, such as tidal tails. Objects whose optical luminosity was dominated by the central QSO component were excluded, since the QSO light prevents reliable morphological classification.

The sample analyzed in \citet{Wojtowicz2023} includes only sources previously examined in an X-ray studied sample by \citet{Gaspari2019}, with the additional requirement of a 1.4\,GHz detection and early-type morphology. In the present work, we relax this criterion by including all early-type galaxies listed in \citet{Bosch2016} with reported 1.4\,GHz fluxes. In their paper, \citet{Bosch2016} compiled black hole mass measurements for a heterogenous sample of 294 galaxies, , of which 141 have early-type morphology. Among these, 34 either lack reported 1.4\,GHz fluxes or have fluxes below the detection limit of the instrument. We include 16 additional sources that were not part of these earlier samples, for which more recent black hole mass estimates are available, and we update the masses of two further sources, namely \emph{PG\,0026+129} and \emph{NGC\,7619}.

\begin{figure*}
    \centering
    \includegraphics[width=0.45\linewidth]{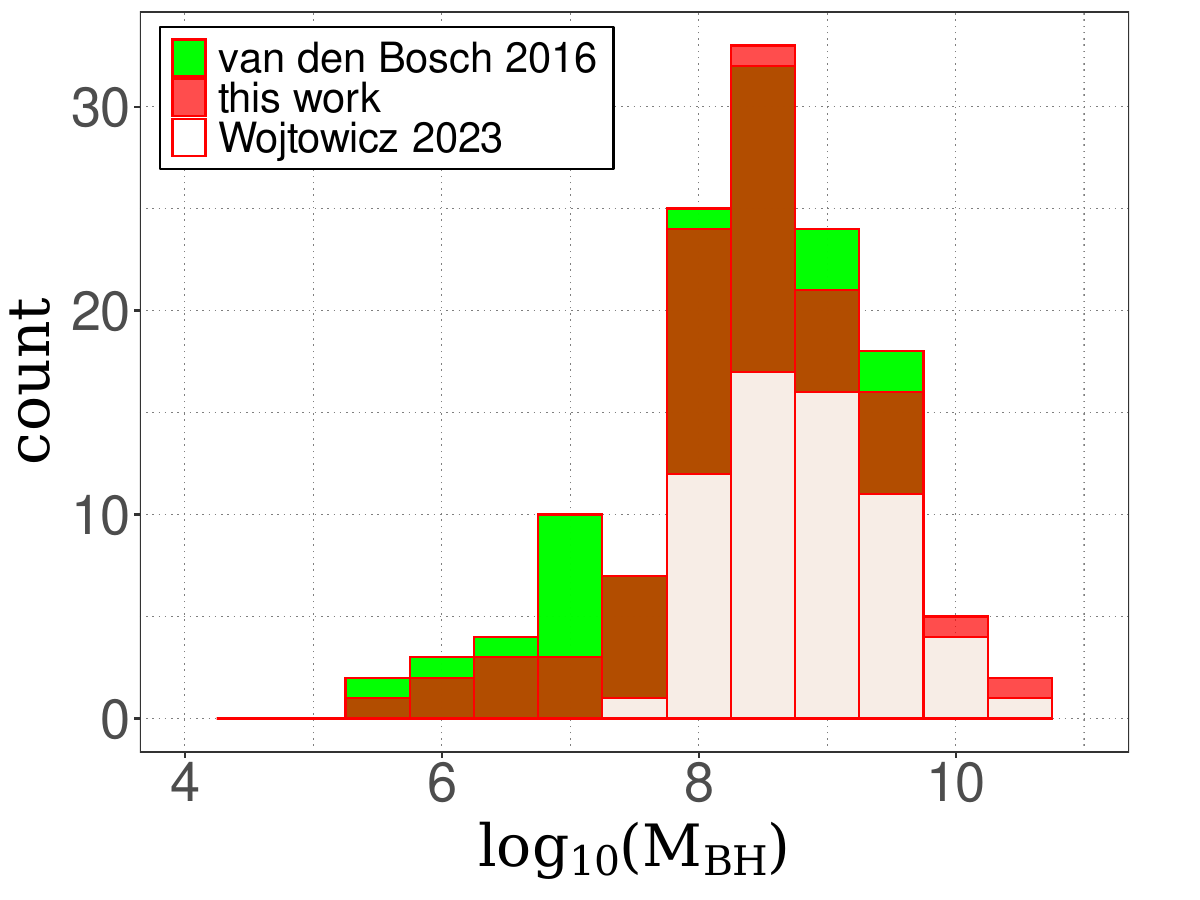}
    \includegraphics[width=0.45\linewidth]{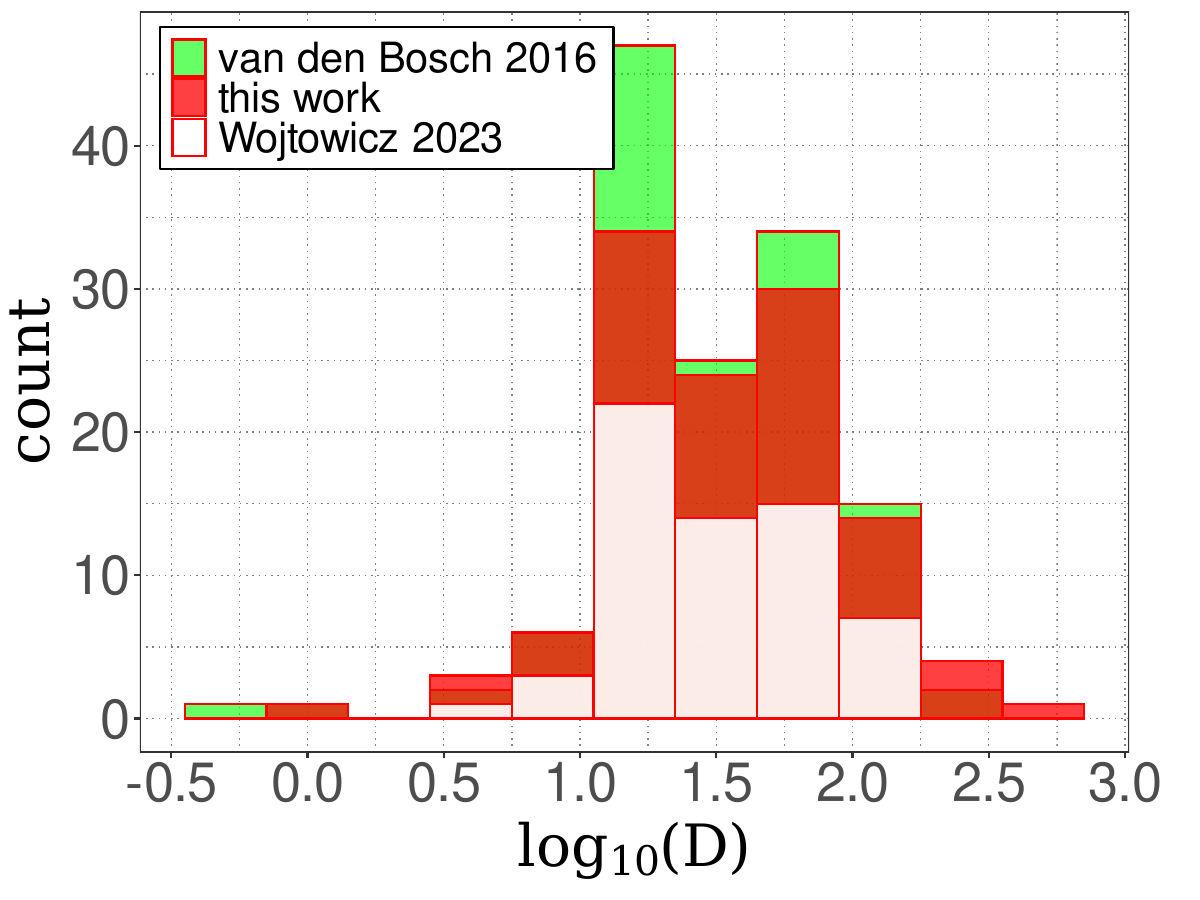}
    \includegraphics[width=0.45\linewidth]{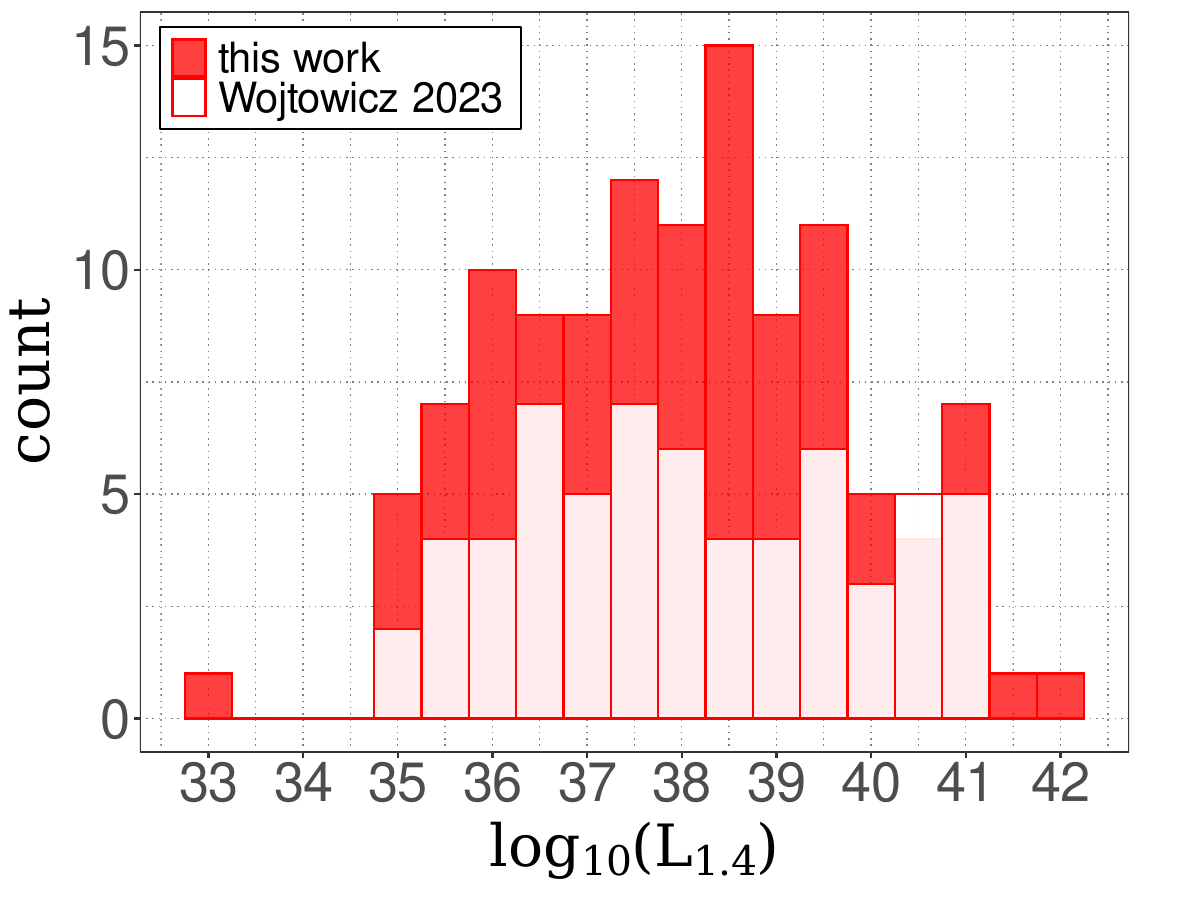}
    \includegraphics[width=0.45\linewidth]{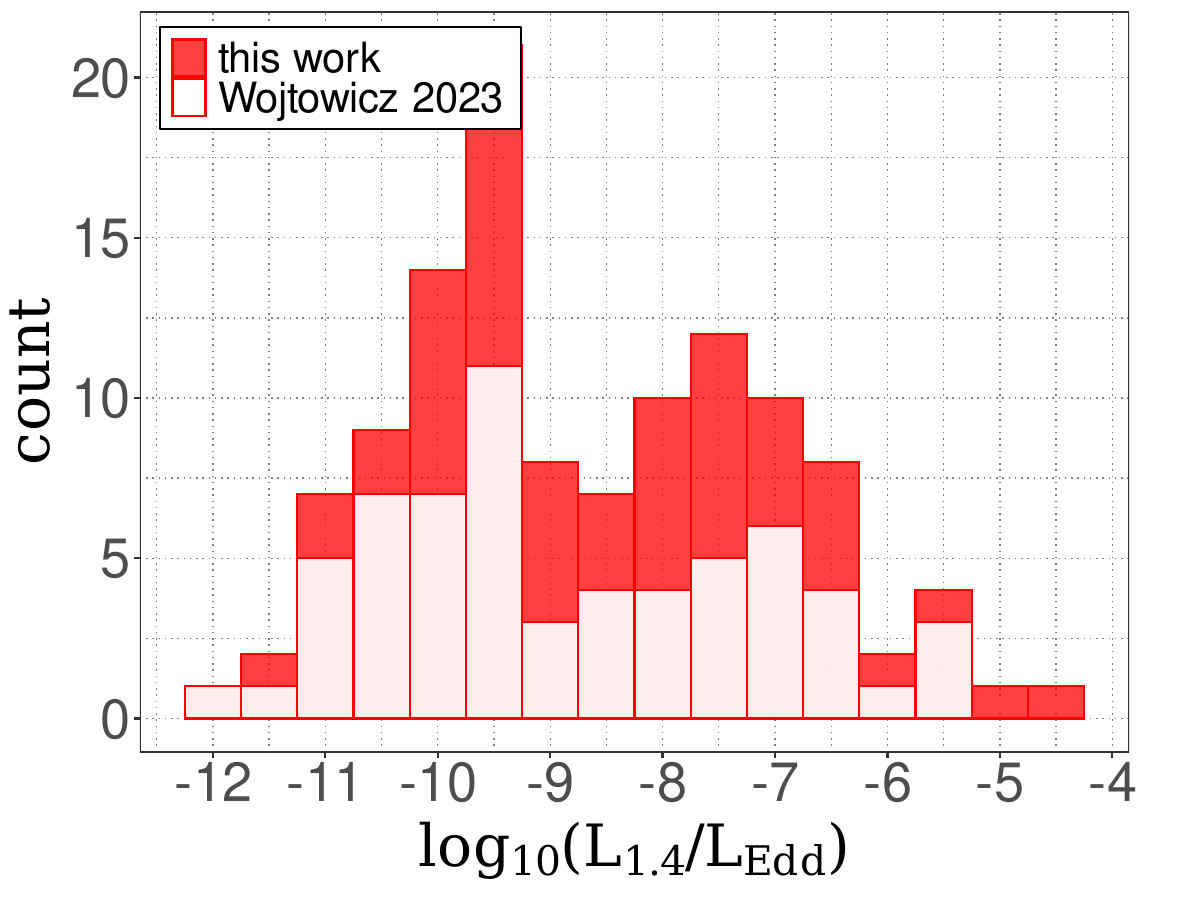}
    \caption{The comparison of source properties presented in this work and in previous studies. The upper-right histogram shows the distribution of $\log_{10} M_{\rm BH}$ for sources presented in this work (red), \citet{Wojtowicz2023} (white), and early-type sources from \citet{Bosch2016} (green). The upper-left panel shows the distribution of source distances (in Mpc), plotted on a logarithmic scale using the same color scheme. The bottom-right and bottom-left histograms show, respectively, the distributions of $\log_{10} L_{1.4}$ and $L_{1.4\,\mathrm{GHz}}/L_{\rm Edd}$ for sources from \citet{Wojtowicz2023} (white) and this work (red).}
    \label{fig:pastandpresent}
\end{figure*}

Figure~\ref{fig:pastandpresent} shows the distributions of black hole mass ($M_{\rm BH}$), distance ($D$), radio luminosity ($L_{1.4,\mathrm{GHz}}$), and Eddington-scaled radio luminosity ($L_{1.4,\mathrm{GHz}}/L_{\rm Edd}$). The top-right panel illustrates the selection effects on the $M_{\rm BH}$ distribution. In \citet{Wojtowicz2023}, the requirement of an hot X-ray halo detection biased the sample towards higher black hole masses (white histogram) and smaller distances, compared to the underlying population of early-type galaxies from the \citet{Bosch2016} parent catalog (green histogram) and the sample presented in this work (red histogram). The lower panels show that the ranges of $L_{1.4,\mathrm{GHz}}$ and $L_{1.4,\mathrm{GHz}}/L_{\rm Edd}$ covered by the samples are however similar.

The full sample consists of 123 early-type galaxies. Six galaxies have only upper-limit estimates for their black hole masses (namely {\it UGC\,1214, NGC\,2685, NGC\,3945, NGC\,4382, UGC\,9799}, and {\it NGC\,4150}), and have been excluded from the analysis presented in the subsequent sections of this paper.  Most of these 123 sources were observed and detected with the NRAO Very Large Array (VLA) at 1.4\,GHz with a 45\,arcsec resolution and are cataloged in \citet{Condon1998}. The corresponding integrated 1.4\,GHz flux densities are also listed in the {\it NASA/IPAC Extragalactic Database (NED)}\footnote{\url{https://ned.ipac.caltech.edu/}}. 

Flux densities for the faintest sources not included in the \citet{Condon1998} cataloged were taken from the \citet{Brown2011} study of a near-infrared-bright (2MASS $K<9$\,mag) sample of early-type galaxies, where the estimates were derived from stacked NVSS images. Three additional sources ({\it NGC\,1023, NGC\,4429}, and {\it NGC\,4459}) that were undetected in \citet{Brown2011} were later successfully detected in 1.4\,GHz VLA observations of early type galaxies from the $ATLAS^{3D}$ survey by \citet{Nyland2017}.

For the brightest extended sources ($S_{\rm 1.4\,GHz} > 0.6$\,Jy), \citet{Brown2011} supplemented the 1.4\,GHz VLA measurements from \citet{Condon1998} with low-resolution ($\approx 12^{\prime}$) single-dish imaging from the 300-ft Green Bank and the 64-m Parkes radio telescopes in order to recover the total extended emission more accurately. Several of the brightest galaxies in our sample ({\it NGC\,7626, NGC\,0741, A\,1836\,BCG, NGC\,3862, NGC\,5128}) were later also re-examined by \citet{Allison2014} using archival 1.4\,GHz NVSS data \citep{Condon1998}. They reported higher and more accurate flux densities than those listed in \citet{Brown2011}, and we adopt the \citet{Allison2014} values for these objects. For two additional bright sources not included in the aforementioned catalogs --- {\it Cygnus\,A} and {\it 3C\,390.3} --- we use the fluxes reported in \citet{Birzan2004} and \citet{White1992}, respectively.

The integrated 1.4\,GHz radio luminosities of the studied galaxies span almost ten orders of magnitude, from $\sim 10^{32}$ to $\sim 10^{42}$\,erg\,s$^{-1}$. Among these sources, 16 are detected only at marginal significance (i.e., $S/N < 3$).

To complement these measurements, we also utilized 3\,GHz VLASS data, which provide 2.5\,arcsec resolution. The 8\,arcsec cutout maps were obtained using the CIRADA server\footnote{\url{http://cutouts.cirada.ca/}}. Radio maps were successfully retrieved for 121 of the 123 objects in our sample; two sources, {\it NGC\,5128} (located at $Dec. < -43^{\circ}$) and {\it Cygnus\,A} (which saturates the VLASS images), are not covered by the survey. Among the VLASS detections, 24 sources show no significant emission at the expected position, 48 exhibit only an unresolved radio core coincident with the host galaxy, and 10 show tentative evidence for small-scale extended structures.

In Appendix\,\ref{sec:radiodata} we show Table\,\ref{tab:radio}, which presents our full sample together with the measured radio fluxes and the corresponding references.

\section{Radio Emission of Early-type Galaxies}
\label{sec:radio}

\subsection{Integrated 1.4\,GHz radio flux}

%stout{In \cite{Wojtowicz2023}, based on \LS{a} sample of 62 early-type galaxies \LS{with directly estimated BH masses}, \LS{$M_{\rm BH}$,} we \LS{identified} a possible bimodality in a distribution of the logarithm of the integrated 1.4\,GHz radio luminosities expressed as a fraction of their Eddington luminosities, $\log L_{\mathrm{1.4\,GHz}}/L_{\mathrm{Edd}}$, \LS{where $\log L_{\mathrm{Edd}}/{\rm erg\,s^{-1}} \simeq 38.1+\log M_{\rm BH}/M_{\odot}$.} 

%That sample composing of 62 sources was insufficient to claim the bi-modality adopting the standard approach, such as \citep{aa2019} excess mass test. 

%In this paper, we present a large sample of 123 sources with significant radio detection at 1.4\,GHz.  \LS{We test whether the bimodality identified in \citet{Wojtowicz2023} remains present in the expanded sample. Six sources with only upper-limit BH mass estimates are excluded from the analysis.} }

\begin{figure*}[h!]
    \centering
    \includegraphics[scale=0.4]{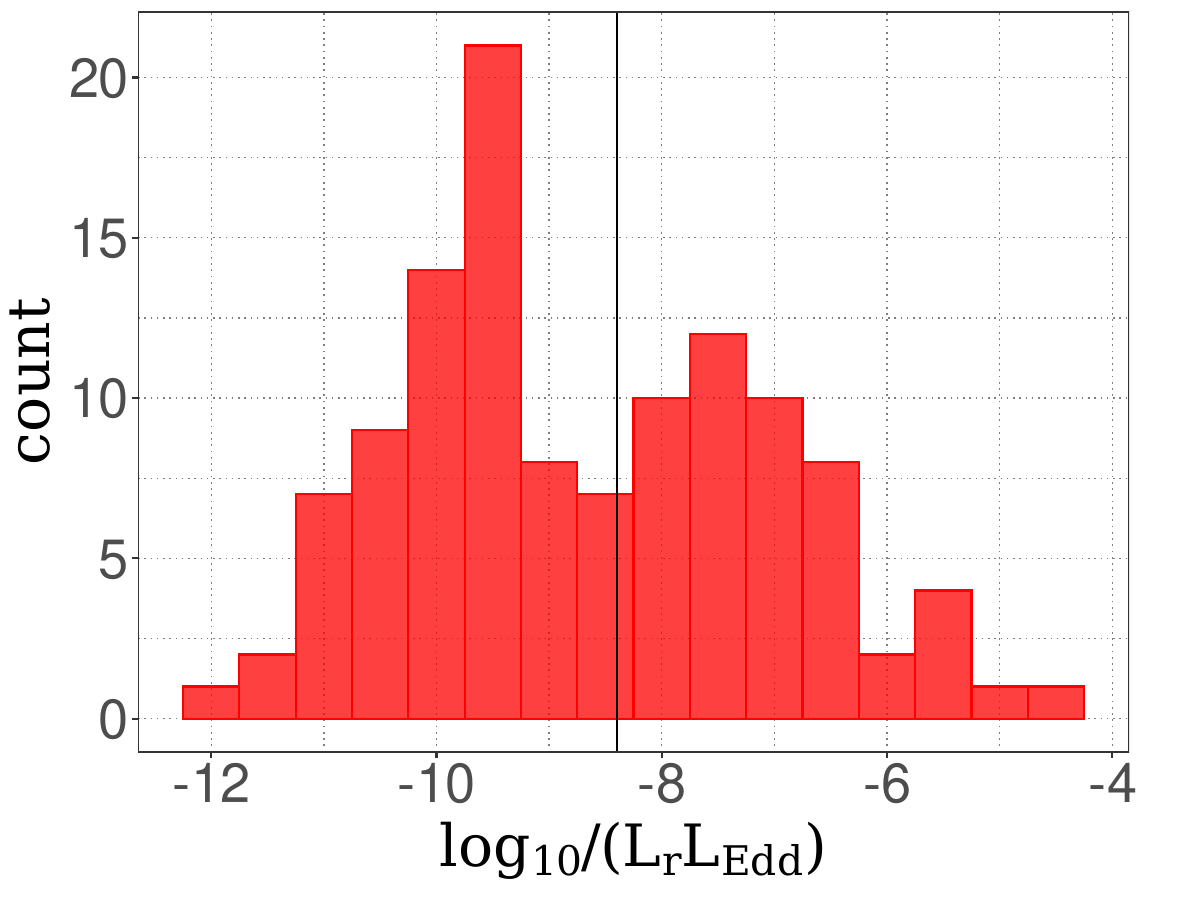}
    \includegraphics[scale=0.4]{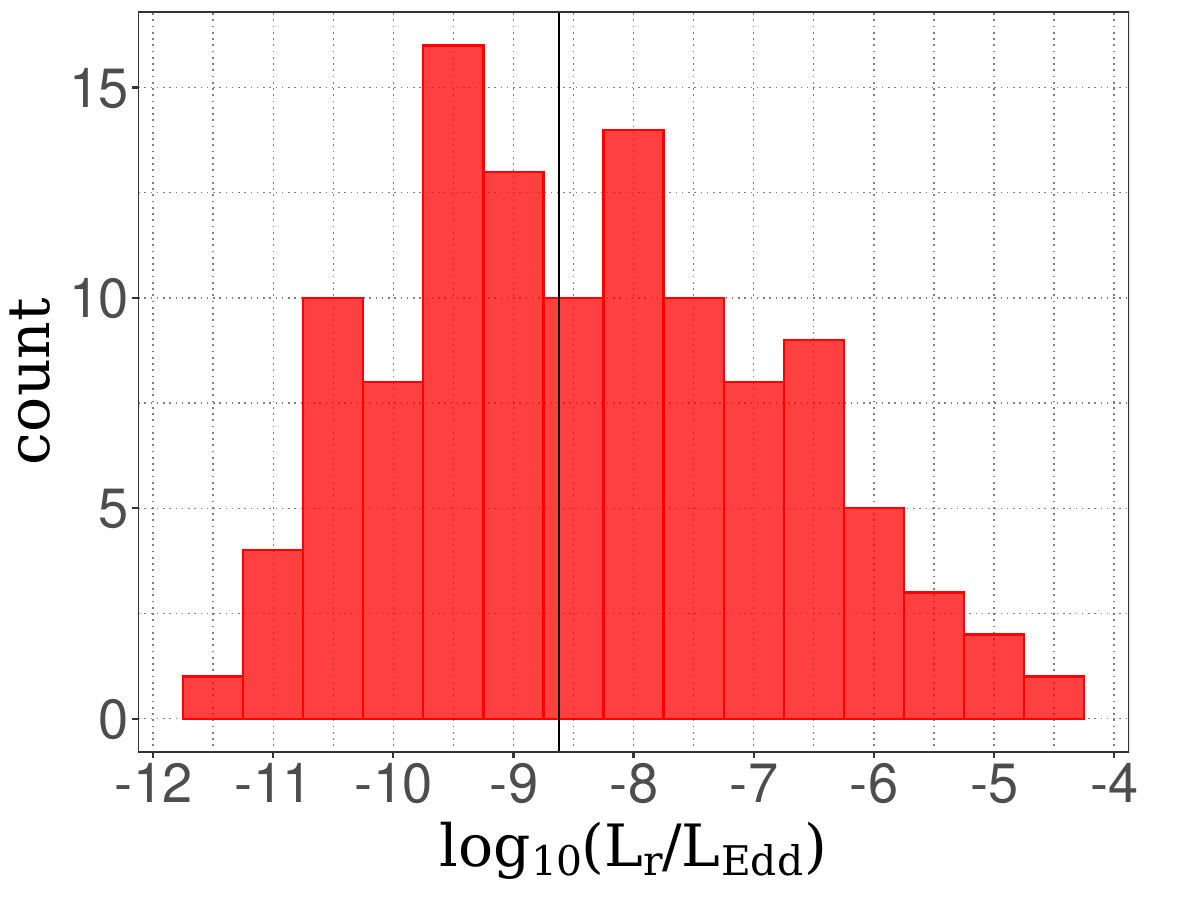}
    \caption{Histogram of integrated 1.4\,GHz radio luminosity expressed in Eddington units, estimated using BH masses determined via direct methods. The position of the antimode at $\log L_{\mathrm{1.4\,GHz}}/L_{\mathrm{Edd}} =-8.6$, identified using the locmodes function, is indicated by a black vertical line. Left: Same as the right panel, but using BH masses estimated indirectly from the $M_{\rm BH}-\sigma_{*}$ relation of \citet{Graham13}; the vertical line indicates the position of the antimode at $\log L_{\mathrm{1.4\,GHz}}/L_{\mathrm{Edd}} = -8.6$.}
    \label{fig:distributions}%
\end{figure*}

In Figure\,\ref{fig:distributions} we show the histogram of $\log L_{\mathrm{1.4\,GHz}}/L_{\mathrm{Edd}}$ for all 117 sources with precise BH masses. The left panel presents the distribution based on BH masses estimated via one of the direct methods, while the right panel presents the corresponding distribution obtained indirectly from stellar velocity dispersion, using the scaling relation of \citet{Graham13}. Notably, the bimodal structure evident in the direct-method distribution disappears when the BH masses are estimated using the scaling relation.

To robustly identify the true number of modes in our integrated $\log L_{\mathrm{1.4\,GHz}}/L_{\mathrm{Edd}}$ distribution, we fit the Gaussian mixture models with number of components ranging from one to five. This method is implemented in the \texttt{GaussianMixtures} function in \texttt{sBIC} package in R-Studio Statistical Software. Based on this analysis and according to the singular Bayesian Information Criterion (sBIC) \citet{Drton2017} we found that the two-mode Gaussian mixture model provides the best description of our data. Next, in order to further test the bimodal hypothesis, we use a maximum likelihood method implemented in the \texttt{mle2} function of \texttt{bbmle} package in R, to fit a mixture of two Gaussians $f(x) = p\times f_1 + (1-p)\times f_2$, where $f_1$ and $f_2$ denote normal distributions, and $p$ is the mixture weight. The parameters of the mixture distribution $f(x)$ obtained through the maximum likelihood method are shown in Table\,\ref{tab:mixture}.

\begin{table}[h]
    \caption{Fitted components of Gaussian mixture model given with function: $f(x) = p\times f_1 + (1-p)\times f_2$, where $f_1\sim N(\mu_1,\sigma_1)$ and $f_2\sim \mathrm{N}(\mu_2,\sigma_2)$ denote normal distributions, and $p$ is the mixture weight. ``VI"-- ``Value Indicated'' in Table\,2 of \citet{Schilling2002}.}
    \centering
    \begin{tabular}{c||c|c|c|c|c|c}
    x & $\mu_{1}$ & $\mu_{2}$ &$\sigma_{1}$ & $\sigma_{2}$ & p & ``VI"\\
    &&&&&&\\
    \hline
    &&&&&&\\
        $\log\frac{L_{\rm 1.4\,GHz}}{L_{\rm Edd}}$ & -9.88 & -7.20 & 0.84 & 1.07 & 0.54 & 1.15   \\
    \end{tabular}
    \label{tab:mixture}
\end{table}

We applied the criterion of \citet{Schilling2002} stating that {\it ``The Mixture Density $f(x) = p\times f_1 + ( 1-p)\times f_2$ Is Bimodal If and Only If $|\mu_2-\mu_1|$ Exceeds $(\sigma_1 + \sigma_2)$ Times the Value Indicated''}, where the ``Value Indicated'' depends on the standard deviation ratio and the weight (see Table\,2 therein). In our case, $\sigma_1/\sigma_2\approx0.78$ and $p\approx0.5$, implying according to \citet{Schilling2002} that our mixture density is bimodal if and only if $|\mu_2-\mu_1|>1.25\times (\sigma_1 + \sigma_2)$. This condition is satisfied for the values of the model parameters obtained using the maximum likelihood method as provided in Table\,\ref{tab:mixture}. This supports the presence of the two modes in the analyzed dataset.

We find the position of the antimode in the $\log{L_{\rm 1.4\,GHz}/L_{\rm Edd}}$ distribution using the R-Studio Statistical Software and \emph{locmodes} function, implemented in the package \emph{multimode} from the CRAN repository. We set the true number of modes \emph{mod0=2} as supported by the analysis presented above. The position of the antimode was found at $\log{L_{\rm 1.4\,GHz}/L_{\rm Edd}}=-8.6$ dividing our sample into two categories ``radio-dim'' objects with $\log{L_{\rm 1.4\,GHz}/L_{\rm Edd}}\leq-8.6$ and ``radio-bright'' if $\log{L_{\rm 1.4\,GHz}/L_{\rm Edd}}>-8.6$.
Here we note that 34 early-type sources from \citet{Bosch2016}, which either lack reported 1.4\,GHz flux measurements or have fluxes below the instrumental detection limit (0.5\,mJy), were not included in this paper. These sources would be classified as radio-dim, with $\log(L_{1.4\,\mathrm{GHz}}/L_{\rm Edd}) \ll -8.6$.

\subsection{VLASS radio core emission at 3\,GHz}

%We extracted the core-radio fluxes with {\it CASA, the Common Astronomy Software Applications package}\footnote{\url{https://casa.nrao.edu/}}, for all sources detected at 3 GHz through careful measurement of the central emission by selecting the $3\,arcsec$ circular regions at the position of radio core. We note that the nuclear flux in the ca se of center brightened extended radio morphology is not always possible, thus in case of these sources the 3 GHz radio core fluxes should be rather interpreted as the upper limits.

We investigate whether the distribution of 3\,GHz core luminosity expressed as the fraction of Eddington luminosity, $\log L_{\rm 3\,GHz}/L_{\rm Edd}$, is also bimodal. We thus utilize the 3\,GHz VLASS radio maps to measure the fluxes associated with core emission, i.e., we measured the fluxes from beam size area centered at the position of host galaxy coordinates. The fluxes were measured through the Analysis task in DS9 software following the Gaussian beam–area formula:
\begin{equation}
F  = S \times \frac{\rm PS^2 \, 4 \ln\!2}{\pi \, \theta_{\rm maj} \, \theta_{\rm min}} \, ,
\end{equation}

\noindent where $F$ is the flux spectral density measured in Jy, $S$ is the sum of pixel values provided in Jy/beam units, PS is the pixel size in arscec$^2$, while $\theta_{\rm maj}$ and $\theta_{\rm min}$ are the beam FWHM axes measured in arsecs. The obtained core flux densities are summarized in Table \ref{tab:radio}.

To investigate the true number of modes in the $\log L_{\rm 3\,GHz}/L_{\rm Edd}$ distribution within our dataset we performed the analogous analysis as we did for 1.4\,GHz integrated radio luminosity. However, a fit of the Gaussian mixture models to the distribution of $\log (L_{\rm 3\,GHz}/L_{\rm Edd})$, with the number of components ranging from one to five, strongly prefer a unimodal Gaussian distribution, thus we found no evidence for a bimodal distribution in core-related $\log (L_{\rm 3\,GHz}/L_{\rm Edd})$.

%\begin{figure}
%    \centering
%    \includegraphics[width=\columnwidth]{bimodality/locmodes_L3Le.pdf}
%    \caption{Probability density function of VLASS extracted radio core luminosity distribution $L_{3GHz}/L_{Edd}$, the position of two present modes was marked by dashed horizontal lines, and the dotted line marks a position of the anti-mode. }
%    \label{fig:my_label}
%\end{figure}

\section{Infrared diagnostics}
\subsection{WISE colours}

\begin{figure}[]
    \centering
    \includegraphics[scale=0.45]{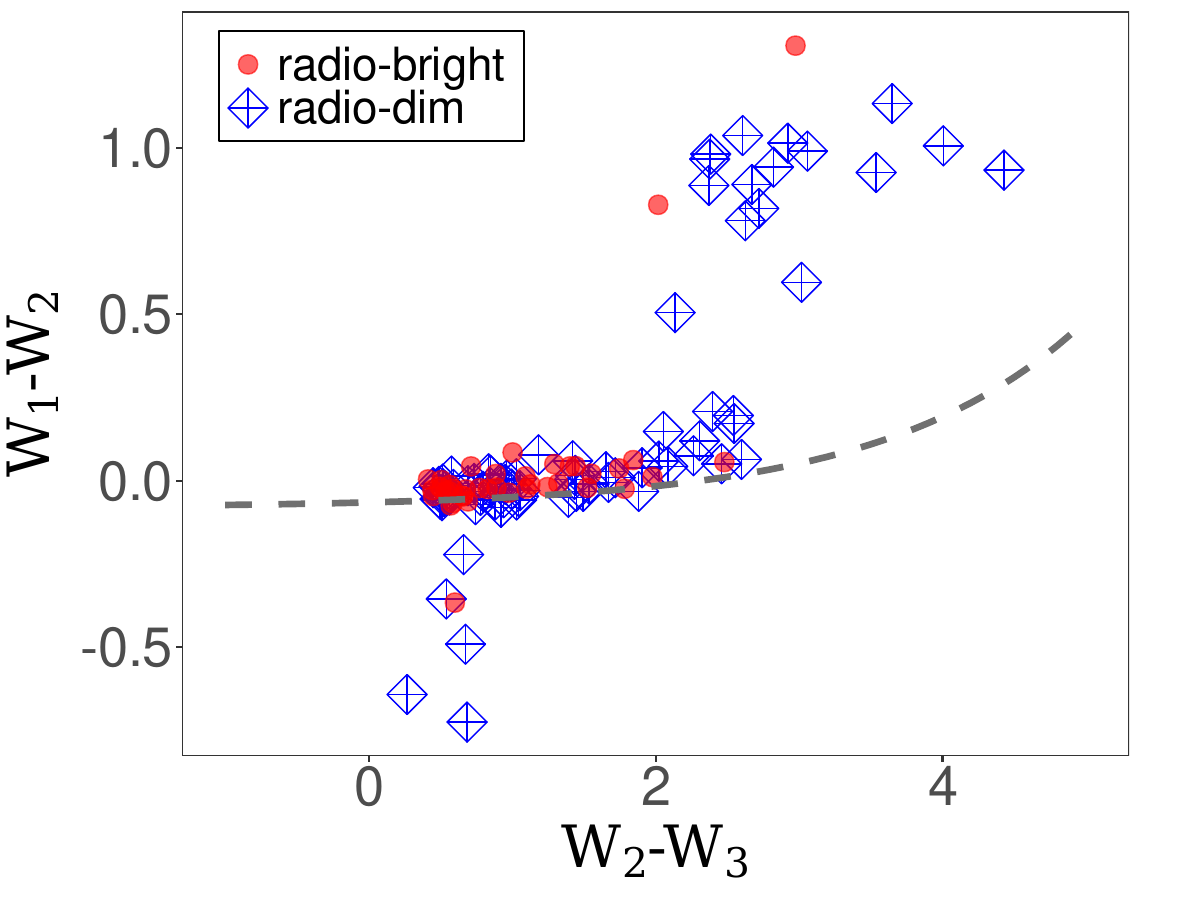}
    \caption{WISE color-color diagram for the analyzed early-type galaxies. Sources belonging to the radio-dim and radio-bright classes are denoted as open blue diamonds and filled red circles respectively. The dashed line indicates the star formation sequence as given in Equation~2.}
    \label{fig:wise}
\end{figure}

WISE magnitudes for the analyzed sources were obtained by cross-matching the sample coordinates with the WISE All-Sky Survey catalog, adopting a matching radius of $12\arcsec$. In Figure\,\ref{fig:wise} we show the radio-faint and radio-bright sources separately on the WISE color-color diagnostic diagram. The corresponding photometric information is summarized in Tables\,\ref{tab:wise}, presented in Appendix\,\ref{sec:wise}.

It is clear that the sample is not homogeneous in terms of nuclear activity. In particular, about 15 radio-bright galaxies show MIR emission dominated by the AGN component, most likely associated with hot circumnuclear dust reprocessing the UV/X-ray radiation from accretion onto the SMBH \citep[see, e.g.,][]{Jarrett11,Stern12,Mateos12}. Two radio-faint galaxies, NGC\,~1194 and NGC\,~5252, also stand out as notable outliers, with NGC\,~1194 exhibiting the most extreme $W1-W2$ color in the sample. The remaining sources closely follow the tight ``star formation sequence'' introduced by \citet{Jarrett19}, which can be approximated by the functional form:
\begin{equation}
({\rm W1-W2}) = 0.015 \times e^{({\rm W2-W3})/1.38} - 0.08
\label{eq:WISE}
\end{equation}
and is indicated in Figure\,\ref{fig:wise} by the dashed curve. Thus, basically all radio-faint sources, as well as the majority of the radio-bright ones, appear to be ISM-dominated in the infrared regime. Consequently, the FIR diagnostics discussed below for assessing the origin of the radio emission are robust, in the sense that they reliably identify the ``radio-excess'' systems.

Note in this context that several of our radio-dim galaxies are characterized by W1$-$W2$ < 0$, indicating that their near-IR emission is dominated by stellar photospheres on the Rayleigh–Jeans tail, with essentially no detectable contribution from warm dust. Such colors place these systems below the star-formation sequence of \citet{Jarrett19} and are characteristic of passive, dust-poor hosts with negligible star formation and no radiatively efficient AGN. In this regime, star formation cannot account for the observed radio emission, and the lack of a mid-IR AGN signature likewise excludes the presence of a persistent accretion disk.

\begin{figure}[]
    \centering
    \includegraphics[scale=0.45]{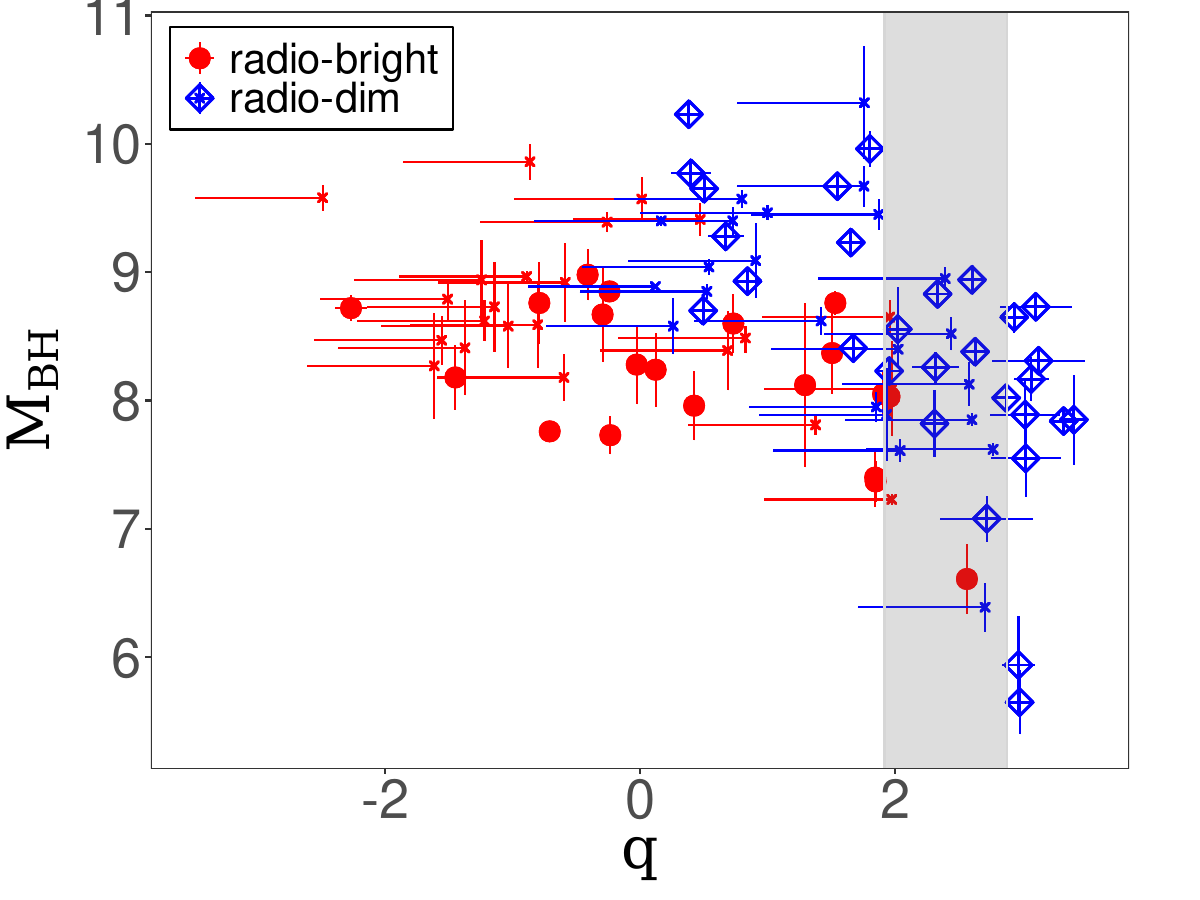}
    \caption{The FIR--radio correlation for the analyzed early-type galaxies. The symbols denote the same classes of sources as in Figure\,\ref{fig:wise}, except for objects with only upper-limit measurements in the far-infrared, which are marked with stars. Grey shaded area denotes the q-values consistent with the range corresponding $q = 2.40 \pm 2\sigma_q $ with $\sigma_q$ = 0.24 \citep{Ivison10}.}
    \label{fig:iras}
\end{figure}

\subsection{Far infrared-radio correlation}
\label{sec:FIRradio}

We compiled $60\,\mu m$ and $100\,\mu m$ far-infrared (FIR) fluxes for all sources in our target list except for 21 sources which lack any FIR measurements. This was achieved by matching the coordinates of our sources with {\it{IRAS Point Source catalog v.2, Serendipitous Survey catalog}} and {\it{Faint Sources catalog}} adopting the matching radius $30$\,arcsec.  If a source was detected in more than one catalog, we adopted the fluxes from our preferred catalog, following the priority order: Point Source Catalog, then the Faint Source Catalog, and finally the Serendipitous Survey Catalog. For sources unidentified in these catalogs we explored NASA/IPAC Extragalactic Database \footnote{\url{https://ned.ipac.caltech.edu/}} to check the availability of the fluxes at 60\,$\mu m$ and 100\,$\mu m$ measured in archival IRAS observations. The final list of the infrared fluxes is given in Tables\,\ref{tab:iras} and \ref{tab:irasupper} presented in the Appendix\,\ref{sec:iras}.
%\subsection{Far infrared-radio correlation}

The FIR--radio correlation is well established at all redshifts, with star-forming galaxies expected to lie within a narrow range of values of the $q$ parameter defined by \citet{Helou85} as
\begin{equation}
    q \equiv \log \frac{F_{\rm FIR}/3.75\times 10^{12}\,{\rm W\,m^{-2}}}{S_{\rm 1.4\,GHz}/{\rm W\,m^{-2}\,Hz^{-1}}} \, ,
\label{eq:q}
\end{equation}
where $F_{\rm FIR}/{\rm W\,m^{-2}} = 1.26 \times 10^{-14} \, [2.58 \times S_{\rm 60\,\mu m}/{\rm Jy} + S_{\rm 100\,\mu m}/{\rm Jy}]$.  In Figure\,\ref{fig:iras} we plot the distribution of this FIR-radio colour in our sample as a function of the BH mass. As before, radio-bright sources with $\log L_{\rm 1.4\,GHz}/L_{\rm Edd}>-8.6$ are denoted in the figure with red color and radio-dim sources with $\log L_{\rm 1.4\,GHz}/L_{\rm Edd}<-8.6$ are marked with blue.  The shaded areas in the plot denote the range corresponding to the median $q=2.40$ with $\pm  2 \sigma_q$ dispersion, where $\sigma_q = 0.24$, as established by \citet{Ivison10} for the galaxies in the GOODS-North field, based on the {\it Herschel} and VLA observations \citep[see also in this context][]{Magnelli2015,Delhaize2017,Giulietti2022}.

It follows from the figure that all sources in the radio-bright sample are indeed over-luminous in the radio relative to their FIR/radio colors. A substantial fraction of the radio-dim sources however exhibit the same behavior, particularly those hosting black holes with masses $\log M_{\rm BH}/M_{\odot}\gtrsim 8.5$. Nevertheless, many radio-dim sources do fall within the star-forming region of the diagram, with some even showing unusually high $q$ values.

\section{Discussion}
\label{sec:discu}

Whereas radio-bright sources exhibit radio emission dominated by synchrotron radiation from a spatially resolved jet, radio-dim systems present a more ambiguous case. At the angular resolution of the 3\,GHz\,VLASS observations (2.5\,arcsec), the radio-dim sources remain unresolved.  In Sect.~\ref{sec:FIRradio}, we showed that a subset of our sources exhibits \textit{FIR/radio} ratios consistent with expectations for star-formation–dominated emission. However, a substantial fraction appears over-luminous in the 1.4\,GHz band, while remaining unresolved in higher resolution 3\,GHz observations.

The physical mechanisms responsible for the compact, low-luminosity radio emission of radio-dim sources therefore remain unclear, suggesting in principle that both star formation and AGN-related processes may contribute. In the following, we discuss the possible origin of the radio emission in our sample, and what mechanisms may give rise to the bimodal distribution of radio luminosities.

\subsection{Star-formation and radio emission}

In general, early-type galaxies exhibit low star-formation rates, typically $\lesssim 0.1\,M_\odot\,\mathrm{yr}^{-1}$, due to the fact that they are generally dust- and gas-poor. The most massive ellipticals can reach values of order $\sim 1\,M_\odot\,\mathrm{yr}^{-1}$ \citep{Kokusho17,Capetti22}. Even such modest levels of star formation can however produce monochromatic radio luminosities up to $\sim 10^{37}\,\mathrm{erg\,s^{-1}}$. Thus, especially in the low-luminosity radio regime $\lesssim 10^{37}\,\mathrm{erg\,s^{-1}}$, the contribution of electrons accelerated in star-forming regions to the total radio emission of early-type systems is never negligible.

 Despite this low level of star formation overall, the spatial distribution and intensity of star-formation activity can be highly diverse in early-type systems depending on the exact amount of cold gas present in the ISM. It was shown for example by \citet{Shapiro10} and \citet{Kuntschner10},  that enhanced star formation in nearby early-type galaxies is confined to fast rotators occurring in two distinct modes: (i) widespread young stellar populations associated with substantial molecular gas reservoirs, and (ii) more compact star-forming structures arranged in disks or rings \citep[see also][]{Young02,Young05,Young11}, and  in some cases, the young stars are concentrated toward the central regions \citet{Kuntschner10}.  

%Using spatially resolved spectroscopy, showed that enhanced star formation in the SAURON sample is confined to fast rotators, occurring in two distinct modes: (i) widespread young stellar populations associated with substantial molecular gas reservoirs, and (ii) more compact star-forming structures arranged in disks or rings \citep[see also][]{Young02,Young05,Young11}. The stellar population maps of 48 early-type galaxies presented by \citet{Kuntschner10} also demonstrate that, in some cases, the young stars are concentrated toward the central regions.\

More evidence on the dual nature of early-type galaxies follows  from the analysis of broad-band spectral energy distribution modeling with CIGALEMC code \citep{Serra2011} by \citet{Amblard2014}. Their results revealed a bimodal distribution in the specific star-formation rates of early-type galaxies, with lenticular galaxies exhibiting, on average, higher specific star-formation rates and dust luminosities than classical elliptical galaxies. This suggests that residual star formation may be more common and extended in at least a subset of early-type systems, particularly those retaining disk-like structures and molecular gas.
%More evidence on the dual nature of early-type galaxies follows  from the analysis of by \citet{Amblard2014}, based on broad-band spectral energy distribution modelling of 221 local early-type galaxies using the CIGALEMC code \citep{Serra2011}, provided more robust estimates of stellar ages and star-formation rates than single-band diagnostics. Their results revealed a bimodal distribution in the specific star-formation rates of early-type galaxies, with lenticulars exhibiting, on average, higher specific star-formation rates and dust luminosities than ellipticals. This suggests that residual star formation may be more common and extended in at least a subset of early-type systems, particularly those retaining disk-like structures and molecular gas.

We note that obtaining reliable star formation rates for the early-type galaxies in our sample is of particular importance for distinguishing the source of radio emission in these systems, particularly at the low-luminosity end. These results will be presented in a dedicated follow-up paper (in preparation).

 In this context, the $q$-factor, defined as the FIR-to-radio luminosity ratio (equation \ref{eq:q}), provides valuable insight into the dominant contribution of young stars to the total radio emission of a galaxy. As discussed in Section\,\ref{sec:FIRradio}, a large fraction of our radio-dim sources, in principle almost all with SMBH masses $\log M_{\rm BH}/M_{\odot}< 8.5$ exhibit $q$-factor values consistent with radio emission being indeed dominated by star-forming regions. However, a notable subset of radio-dim sources, and in particular all radio-dim systems with SMBH masses $\log M_{\rm BH}/M_{\odot}> 9.0$, shows evidence of radio excess, indicating the presence of an additional emission mechanism. Thus, despite the fact that there is compelling evidence for the bimodal nature of the star-formation rate in early-type galaxies, this alone cannot fully explain the observed bimodal distribution in $\log L_{\rm 1.4\,GHz}/L_{\rm Edd}$, which instead appears to be linked to the masses of the central SMBHs.

%%%%%%Reliable estimates of the “pure” radio emission associated with the SMBH would be particularly valuable in cases where the radio emission is unresolved at the available arcsecond-scale resolution, allowing a closer examination of different radio-triggering mechanisms and potentially distinguishing between different SMBH feeding processes.

\subsection{The role of star--SMBH interaction in triggering the radio activity.}

 Launching extended, powerful jets spanning more than hundreds of kpc, as observed in some sources, requires a sustained and long-lasting supply of matter to the accretion flow. On the other hand, if the matter supply is more sporadic and arrives in small parcels, the jets will not develop large-scale structures and will instead remain confined to the host galaxy environment.

 Indeed, such a mechanism was proposed by \citet{Readhead2024} and further developed by \citet{Sullivan2024} to explain the emission properties of radio galaxies belonging to the Compact Symmetric Object (CSO) class. These systems show a sharp cutoff in the distribution of the linear sizes of their radio structures at $\sim 500$\,pc \citep{Kiehlmann2024a, Kiehlmann2024b}, as well as high separation velocities of their terminal hotspots that imply short jet lifetimes. It has been proposed that the radio jets in such sources may be triggered by a single tidal disruption event (TDE), for example, the disruption of a giant-branch star in the vicinity of the central supermassive black hole. 

As argued by \citet[][see Section\,5 therein]{Sullivan2024}, if the disrupted star carries an average magnetic field of $\gtrsim 100$\,G, and passes only within $\sim 100$ gravitational radii from a SMBH, the event could still supply sufficient magnetic flux to seed jet formation and power outflows with luminosities of order $\sim 10^{33}\,\rm erg\,s^{-1}$ via the \citet{BZ1977} mechanism; subsequent evolution of the infalling debris may possibly further amplify the magnetic field near the SMBH, providing additional fuel supply capable of powering even the most luminous CSO sources. In our case, i.e., radio-dim systems unresolved at radio frequencies (arcsecond scales), with black-hole masses of $M_{\rm BH} \sim (10^9-10^{10}) \, M_{\odot}$, integrated 1.4-GHz luminosities of $L_{\rm 1.4\,GHz} \sim (10^{-12}-10^{-9}) L_{\rm Edd} \sim 10^{35}-10^{39}$\,erg\,s$^{-1}$ --- which is far below the radio luminosities of the classic CSOs discussed by \citet{Readhead2024} and \citet{Sullivan2024} --- and $q$-parameters indicating radio emission in excess of that expected from ongoing star formation, the required energetic output is modest enough to be supplied by such TDE-driven, short-lived jet episodes.

%A particularly interesting class of unresolved sources at arcsecond resolution are the compact symmetric objects (CSOs). %It is clear that majority of sources unresolved in scales of \textcolor{red}{(5 \,arcsec)} are associated with high-excitation galaxies, in radio-spectrsocopcially limited sample we see that substantial number of unresolved sources are found in gas-poor galaxies.
%Studies of 79 bona fide CSOs reveal a sharp radio size cutoff at $\sim 500$~pc \citep{Kiehlmann2024a, Kiehlmann2024b}, and the high separation speeds of their hot spots imply short lifetimes. Based on their brief lifespans and diverse morphologies, \citet{Readhead2024} suggested a link between CSOs and tidal disruption events (TDEs). Building on this idea, \citet{Sullivan2024} proposed that small FR,II-type jets can be triggered by a single TDE. For example, the disruption of a giant branch star ($R = 100,R_\odot$, $M = 1,M_\odot$) by a SMBH of $\log M_{\rm BH} = 8~M_\odot$, with a magnetic flux $B \gtrsim 20^2$~G passing at $\sim 100$ gravitational radii, can generate a seed magnetic field sufficient to power jets of $\sim 10^{33},\rm erg,s^{-1}$ via the \citet{BZ1977} mechanism. Moreover, the gravitational energy of the infalling debris can further amplify the magnetic field near the SMBH, providing enough energy to account for the initial jet energetics of CSOs. While stronger seed fields are required in ballistic jet models to explain the full observed energetics, this mechanism remains a plausible explanation for the compact radio emission observed in some of our radio-dim sources.

\subsection{Merger history of host galaxies}
\label{sec:mergers}

Some information on the merger history of the host galaxy can be assessed through analysis of their (starlight) luminosity profiles. In particular, some luminous early-type galaxies exhibit partially depleted stellar cores, observed as a flattening of the luminosity profile in the inner region \citep{Graham2003}. A widely favored explanation for this feature involves the presence of a supermassive binary black hole, which ejects stars into radial orbits via three-body interactions \citep[][]{Begelman1980, Ebisuzaki1991, Quinlan1996, Yu2002, MM2005}. In our sample only 12 of the 35 radio-dim galaxies with well-characterized luminosity profiles show evidence of core depletion, while seven of eleven radio-bright galaxies with studied luminosity profiles possess depleted cores \citep[see:][]{Savorgnan2016,Davis2019,Sahu2019a}.

%Many luminous early-type galaxies exhibit partially depleted stellar cores, observed as a flattening of the inner luminosity profile \citep{Graham2003}. A widely favored explanation involves the presence of a supermassive binary black hole: following a massive dry merger, the black holes sink to the center and form a binary system that ejects stars on radial orbits via three-body interactions, producing a depleted core \citep[][]{Begelman1980, Ebisuzaki1991, Quinlan1996, Yu2002, MM2005}. In our sample, 15 of the 40 radio-dim galaxies with well-characterized luminosity profiles show evidence of core depletion, while five of eight radio-bright galaxies with available profiles possess depleted cores \citep[see:][]{Savorgnan2016,Davis2019,Sahu2019a}.

Another piece of information on the violent past of some early-type galaxies can also be obtained from their internal kinematics. The ATLAS$^{\rm 3D}$ survey of a volume-limited sample of local early-type galaxies found that most of them exhibit significant rotation along their major axes \citep{Cappellari2011a, Cappellari2011b, Emsellem2011, Krajanovic2011}. Of the 35 galaxies from our sample included in ATLAS$^{\rm 3D}$, 29 are radio-dim and only six are radio-bright. Among the radio-bright subsample, 50\% of the galaxies are classified as fast rotators, while 22 out of 29 radio-dim galaxies (76\%) fall into this category.

Although radio-bright sources are generally under-represented in such surveys, these statistics suggest that radio-dim galaxies are preferentially associated with fast-rotating systems and are more likely to lack depleted cores, pointing to their less violent merger history.

\subsection{$M_{BH}-\sigma_*$ scaling relation}

The observed strong correlation between dynamically measured black hole masses, $M_{\rm BH}$, and the stellar velocity dispersion within the effective radius of host galaxies, $\sigma_*$, is considered a clear manifestation of the co-evolution of central SMBHs and their hosts. Since its discovery \citep{Gebhardt2000,Ferrarese2000}, significant effort has been devoted to calibrating and understanding this correlation \citep[e.g.,][]{Bosch2016,Sahu2019a,Baldassare2020ApJ}.

Recently, \citet{Smethurst2023} used the output from the Horizon-AGN simulation to show that galaxies experiencing more than three mergers since redshift $z = 2$ exhibit a shallower correlation slope, whereas isolated galaxies are less tightly correlated and on average have steeper slopes. Similarly, \citet{Beckmann2023} reported that galaxies undergoing recent mergers tend to have lower spin values compared with those evolving in isolation.

Motivated by these findings, we examine the role of mergers and BH spin in shaping the observed radio-dim/radio-bright bimodality by analyzing the slopes of the respective $M_{\rm BH}-\sigma_*$ correlations.

\begin{figure*}[b!]
    \centering
    \includegraphics[scale=0.4]{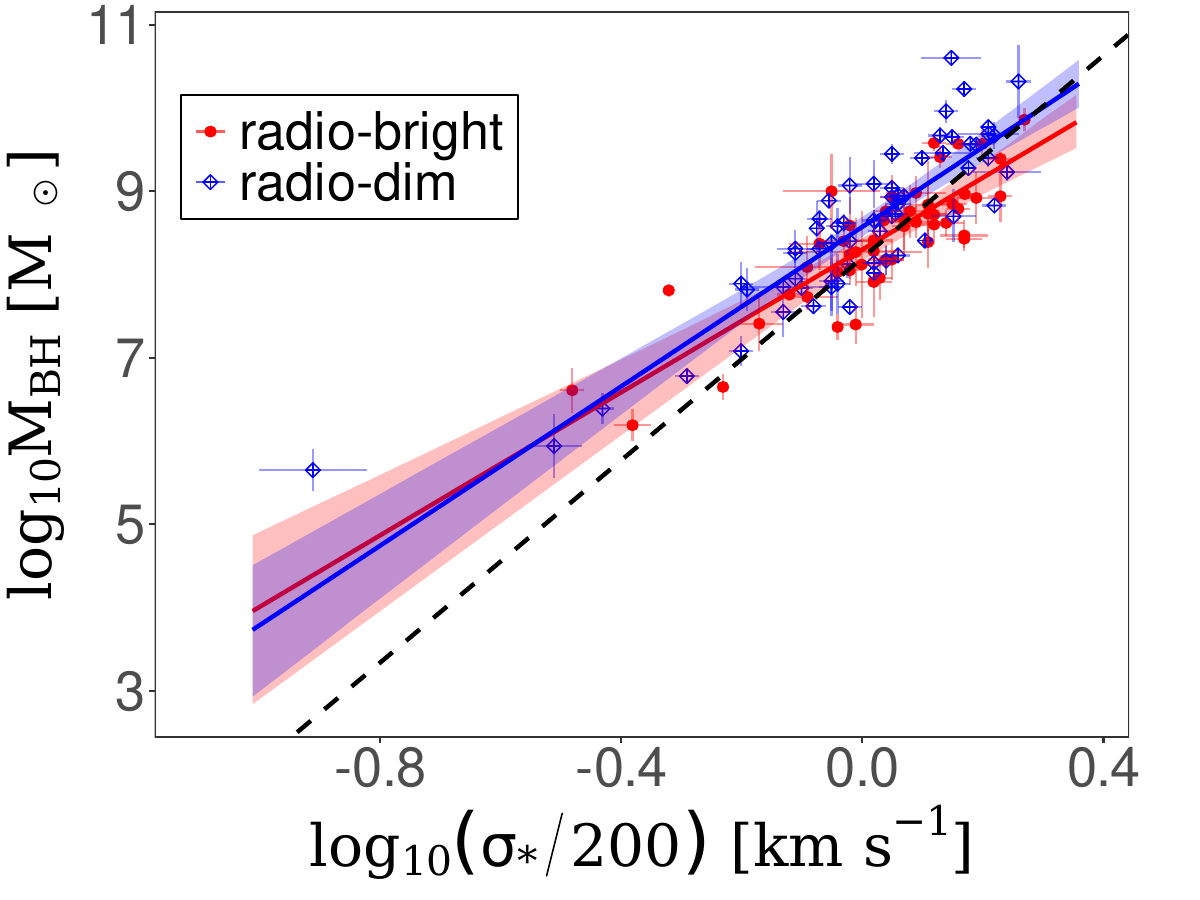}
    \includegraphics[scale=0.45]{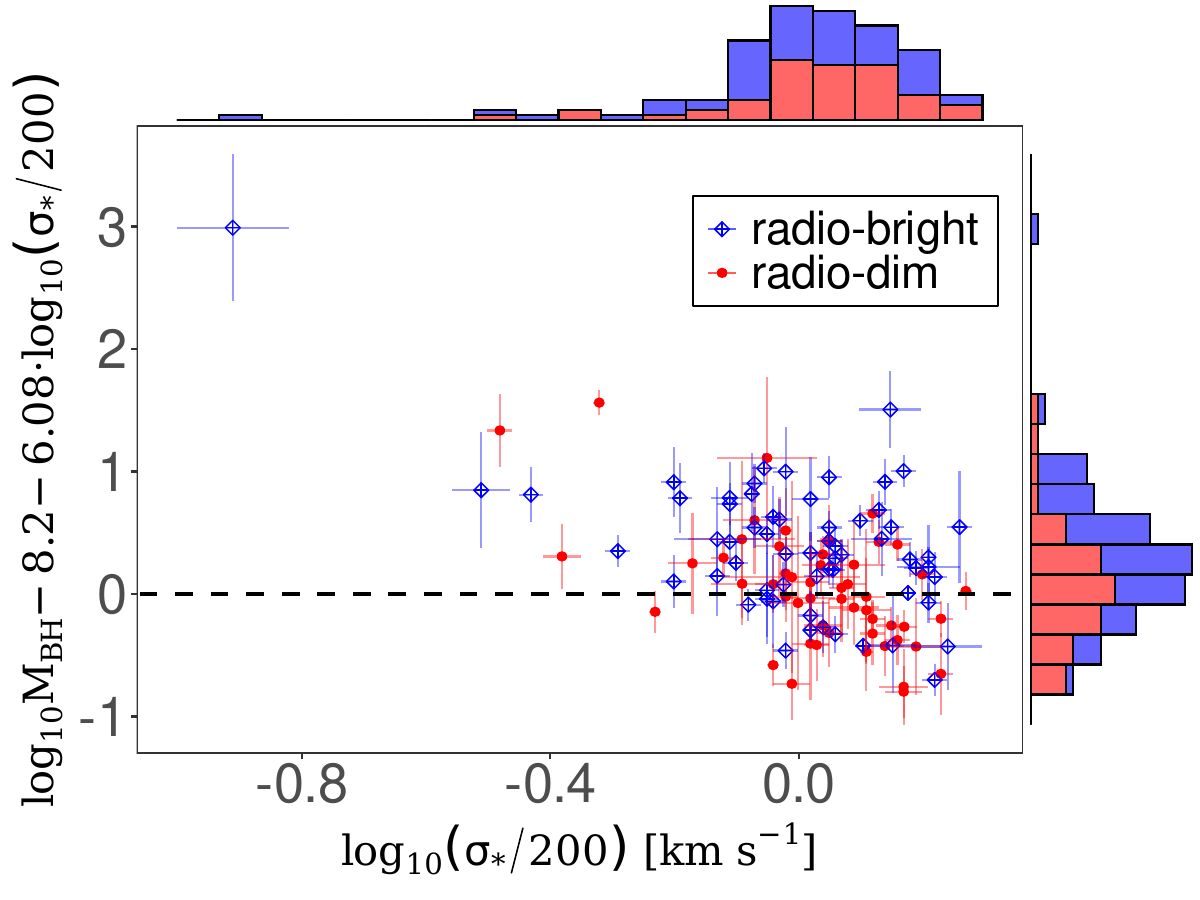}
    \caption{Left panel shows the relation between black hole mass ($\log_{10} M_{\rm BH}$) and stellar velocity dispersion ($\log_{10}(\sigma_*/200\,\mathrm{km\,s^{-1}})$) for the early-type galaxies in our sample. Radio-dim sources are shown as blue diamonds and radio-bright sources as red circles, and their corresponding regression lines plotted as solid lines with shaded areas indicating 95\% credibility intervals. The black dashed line shows the regression obtained by \citet{Graham13} for their full sample of 77 galaxies. Right panel shows the scatter of the radio-dim and radio-bright sources from the mean regression line obtained by \citet{Graham13}, as a function of velocity dispersion. The projected histograms of the source distributions along both axes are presented with the same color scheme for the two radio sub-types.}
    \label{fig:mbh-sigma}
\end{figure*}

In Figure~\ref{fig:mbh-sigma}, we plot the $M_{\rm BH}-\sigma_*$ relation for our sample of early-type galaxies separated by radio-dim (blue) and radio-bright (red). Interestingly, the bimodality in our dataset is also reflected in the slightly different correlations of the two classes of objects in the $M_{\rm BH}-\sigma_*$ diagram. To quantify this difference, we perform a univariate linear regression analysis using the APEMoST algorithm\footnote{Automated Parameter Estimation and Model Selection Toolkit; \url{http://apemost.sourceforge.net/}, 2011 February}\citep{Gruberbauer2009}, with the response variable $Y = \log_{10} M_{\rm BH}$ and the predictor variable $X = \log_{10} (\sigma_\star / 200~{\rm km~s^{-1}})$. We assume a linear trend,
\begin{equation}
Y = a + b \cdot X + \epsilon,
\end{equation}
where the noise term $\epsilon$ is normally distributed, $\epsilon \sim \mathcal{N}(0,\sigma)$. The standard deviation of the noise is expressed as
\begin{equation}
\sigma = \sqrt{\sum_{i=1}^{N} \sigma_{\rm int}^2 + \sigma_{Y_i}^2 + (b , \sigma_{X_i})^2} \, ,
\end{equation}
with $N = 46$ and $N = 71$ for the radio-bright and radio-dim samples, respectively. Here, $\sigma_{\rm int}$ represents the intrinsic scatter, and $\sigma_{X_i}$ and $\sigma_{Y_i}$ denote the respective measurement uncertainties associated with $X_i$ and $Y_i$. We aim to determine the multidimensional probability density function (PDF) of the parameter set $\theta = {a, b, \sigma_{\rm int}}$.

Following \citet{Ostorero17} and \citet{Wojtowicz21}, we perform $2 \times 10^6$ MCMC iterations with 20 chains to ensure robust sampling of the parameter space. The resulting regression parameters are presented in Table~\ref{tab:bayes}.

\begin{comment}We assume flat uninformative priors for the $a$ and $b$ parameters, with the parameter space boundaries set to  $[-100,100]$. The prior of the intrinsic spread $\sigma_{\rm int}$, which is always a positive \TC{value}, is given by the PDF that describes a variate with mean $r/\mu$ and variance $r/\mu^2$, namely,
\begin{equation}
P\!\left(\sigma_{\rm int}|\mathcal{M}\right)=\frac{\mu^r}{\Gamma(r)} \, x^{r-1} \, \exp(-\mu x) \, ,
\end{equation}
where $x= 1/\sigma_{\rm int}$, and $\Gamma(r)$ is the Euler Gamma function; in our calculations, we set $r=\mu=10^{-5}$ and the variability interval boundaries $[0.01,1000]$. The random number generator was set with bash command \texttt{GSL\_RNG\_TYPE="taus"} and the initial seed of the random number generator was set with \texttt{GSL\_RNG\_SEED=\$RANDOM}. 
The medians of the parameter PDFs that follow from this regression analysis are reported in Table~\ref{tab:bayes} $\tilde{\theta} =\{\tilde{a},\tilde{b};\tilde{\sigma}_{\rm int}\}$, with the associated uncertainties.
\end{comment}

\begin{table}[h!]
    \centering
        \caption{Summary of the Bayesian linear regression analysis for the radio-dim and radio-bright sub-samples.}
    \begin{tabular}{c||c|c|c}
         Sub-sample & a  & b & $\sigma_{int}$  \\
         & & & \\
         \hline
           & & & \\
        Radio-dim &  {$8.57\pm{0.06}$} &  {$4.79\pm0.38$} &  {$0.43\pm0.05$}\\
        Radio-bright & {$8.31\pm0.07$} & {$4.29^{+0.46}_{-0.43}$}& {$0.38^{+0.06}_{-0.05}$}\\
    \end{tabular}

    \label{tab:bayes}
\end{table}

Our analysis hints that the radio-dim and radio-bright sources may occupy different loci on the $M_{\rm BH}$–$\sigma_{*}$ plane, though the substantial scatter within each subset limits the statistical significance.
What we see, in particular, is that radio-bright objects from our sample follow shallower correlation than the one observed in radio-dim objects. Looking at the right panel of Figure\,\ref{fig:mbh-sigma}, we see that, to first order, the radio-bright galaxies follow the mean $M_{\rm BH}$–$\sigma_{*}$ relation of \citet{Graham13}, whereas the radio-dim sources tend to lie slightly above that scaling. This suggests that the SMBHs in the radio-dim systems are, on average, overmassive relative to the stellar velocity dispersion of their hosts compared with the general galaxy population used by \citet{Graham13}.

For context, the \citet{Graham13} relation is based on a sample of 77 nearby galaxies with reliably measured SMBH masses and host-galaxy velocity dispersions. Their best-fitting power-law slope for this full sample, as well as for a refined subsample of 72 galaxies comprising 24 core-Sersic and 48 Sersic spheroids, is $b = 6.08\pm0.41$ and $6.08\pm0.31$, respectively. The sample spans a wide range of morphological types, including ellipticals, lenticulars, and spiral bulges, making it a broadly representative reference set for `normal' galaxies.

%the correlation was studied for type 1 and type 2 AGN selected from the BAT survey

It is interesting to note in this context the recent findings of \citet{Gliozzi2024}, who report that AGN in their sample produce a shallower $M_{\rm BH}$–$\sigma_{*}$ correlation than that obtained for quiescent galaxies \citep{Kormendy&Ho2013}. A similar behavior is seen in our dataset: the radio-dim galaxies presented in this paper yield a steeper correlation than the mean regression line of \citet{Graham13}, whereas the radio-bright objects produce a comparatively shallower slope.

%In the later case following our discussion from this section the radio-bright sources are found in the system which underwent through major mergers while radio-dim sources are more likely to be found in more isolated systems. 

%{\textcolor{orange}{Alternatively one can argue that if the radio-bright galaxy evolved as the results of major merger than higher velocity dispersion should be expected. We showed in section \ref{sec:mergers} that in framework of evidences provided in this paper we cannot robustly confirm if radio-bright sources are more likely to be the product of major mergers.}}

\section{Conclusions}
\label{sec:concl}

We investigate the distribution of Eddington-scaled radio luminosities among 117 nearby early-type galaxies with directly measured black hole masses. Using integrated 1.4\,GHz flux densities and with approximately twice larger sample than analyzed in \citet{Wojtowicz2023}, we confirm a bimodality in the $L_{\rm 1.4\,GHz}/L_{\mathrm{Edd}}$ distribution, with peaks separated at $\log (L_{\rm 1.4\,GHz}/L_{\mathrm{Edd}})\approx -8.6$, dividing our sample into radio-dim ($\log (L_{\rm 1.4\,GHz}/L_{\mathrm{Edd}})\leq -8.6$) and radio-bright ($\log (L_{\rm 1.4\,GHz}/L_{\mathrm{Edd}})>-8.6$) objects. This structure is however only recovered when using dynamical black hole masses, and is washed out when masses from the $M_{\rm BH}-\sigma_\star$ relation are adopted, or only 3\,GHz core luminosity is included.

High-resolution VLASS 3\,GHz imaging shows that the radio-bright subset consists almost exclusively of galaxies hosting spatially extended jets, while radio-dim systems typically display compact nuclear emission. Many radio-dim galaxies, in particular those associated with the most massive black holes, nevertheless show a radio-excess relative to the far-infrared--radio correlation, indicating that the radio emission cannot be produced solely by star-forming processes, but must instead be dominated by mechanisms associated with SMBH activity.

The two radio populations appear to differ in their host-galaxy structural properties: the radio-bright systems are predominantly slow rotators with depleted stellar cores, whereas the radio-dim systems are mostly fast rotators. This distinction is further supported by their separation on the $M_{\rm BH}$–$\sigma_{*}$ plane, where the radio-dim sources exhibit a steeper regression slope than that obtained for the radio-bright objects. Moreover, when compared with the general population of normal galaxies of various morphological types studied by \citet{Graham13}, the radio-dim systems appear, on average, to host overmassive SMBHs relative to their stellar velocity dispersions.

In our view, this points toward a scenario in which the radio-dim systems represent the low-energy tail of the radio-luminosity distribution normalized by SMBH mass in early-type galaxies. Their radio output likely reflects modest, intermittent supplies of magnetized gas delivered to the accretion flow through stochastic processes, most plausibly the tidal disruption of giant-branch stars passing through the immediate vicinity of the SMBH. Because the mass supply in such events is relatively uniform, the resulting accretion episodes produce broadly similar levels of radio emission across early-type systems. The observed bimodality in the Eddington-normalized radio-luminosity distribution would then arise primarily from the fact that a subset of fast rotators host overmassive black holes.

If correct, this scenario offers a coherent physical explanation for the bimodality in Eddington-normalized radio power at the lowest, rarely probed luminosity range in early-type galaxies, linking SMBH radio output to stellar kinematic class and black-hole overmassiveness, and identifying stochastic stellar-disruption events as a universal fueling channel for low-level radio activity in these systems.

\begin{acknowledgements}
 A.~W. and N.~W. were supported by the GACR grant 21-13491X. Research by C.~C.~C. at the Naval Research Laboratory is supported by NASA DPR S-15633-Y.
\end{acknowledgements}

\appendix
\onecolumn
\section{WISE data}
\label{sec:wise}
\begin{table*}[h!]
    \caption{WISE colours of radio-bright (left part of the table) and radio-dim objects (right part of the table) obtained based on magnitudes measured with profile-fitting photometry in the WISE All-Sky Survey catalogue.}
\tiny
    \centering
    \begin{tabular}{|lcc|lcc|}
    \hline
    \multicolumn{3}{|c|}{radio-bright galaxies}&\multicolumn{3}{|c|}{radio-dim galaxies}\\
    \hline    \hline
Name&W1-W2&W2-W3&Name& W1-W2&W2-W3\\
&\tiny{[mag]}&\tiny{[mag]}&&\tiny{[mag]}&\tiny{[mag]}\\
(1)&(2)&(3)&(1)&(2)&(3)\\
    \hline
PG\,0026+129&0.966&2.377&ESO\,444-G25&-0.074&0.571\\
3C\,120&1.014&2.921&Holm15A&-0.045&0.652\\
3C\,390.3&1.037&2.607&MRK\,1216&-0.062&0.692\\
A1836BCG&-0.072&0.745&NGC\,0221&-0.642&0.268\\
Ark120&0.887&2.371&NGC\,0404&0.056&2.479\\
Cygnus\,A&0.933&4.428&NGC\,0524&0.004&1.435\\
IC\,1459&-0.02&0.847&NGC\,0584&0.001&0.564\\
MRK\,0279&0.926&3.537&NGC\,1023&-0.048&0.524\\
MRK\,0509&0.89&2.671&NGC\,1194&1.307&2.975\\
MRK\,1310&0.596&3.017&NGC\,1332&-0.025&0.995\\
MRK\,1383&0.942&2.822&NGC\,1358&-0.005&1.671\\
MRK\,335&0.981&2.384&NGC\,1380&0.016&1.44\\
MRK\,771&0.99&3.058&NGC\,1399&-0.222&0.662\\
NGC\,0193&-0.08&0.922&NGC\,1407&-0.032&0.538\\
NGC\,0315&0.039&1.906&NGC\,1453&0.002&0.508\\
NGC\,0383&0.027&1.653&NGC\,1497&0.208&2.397\\
NGC\,0541&-0.058&1.031&NGC\,1550&-0.067&0.586\\
NGC\,0741&-0.021&0.45&NGC\,1600&-0.028&0.444\\
NGC\,1052&0.196&2.542&NGC\,2179&-0.024&1.784\\
NGC\,1275&1.006&4.005&NGC\,2693&-0.011&1.127\\
NGC\,2110&0.781&2.625&NGC\,2787&-0.008&1.322\\
NGC\,2329&0.006&1.03&NGC\,2911&-0.021&1.524\\
NGC\,2892&-0.026&0.591&NGC\,2974&-0.019&1.246\\
NGC\,3078&0.021&0.836&NGC\,3091&-0.031&0.536\\
NGC\,3557&-0.015&0.691&NGC\,3115&-0.725&0.686\\
NGC\,3801&-0.032&1.882&NGC\,3245&0.011&1.978\\
NGC\,3862&0.06&1.422&NGC\,3379&-0.355&0.541\\
NGC\,4151&0.818&2.719&NGC\,3414&-0.036&0.977\\
NGC\,4261&-0.03&1.055&NGC\,3489&0.043&1.402\\
NGC\,4278&0.002&0.962&NGC\,3585&-0.008&0.735\\
NGC\,4335&-0.032&1.447&NGC\,3607&0.037&1.746\\
NGC\,4374&0.016&0.85&NGC\,3608&-0.024&0.517\\
NGC\,4486&0.078&1.184&NGC\,3640&-0.018&0.495\\
NGC\,5127&-0.051&0.941&NGC\,3665&0.064&2.599\\
NGC\,5128&0.505&2.136&NGC\,3706&-0.013&0.567\\
NGC\,5283&0.12&2.307&NGC\,3842&-0.046&0.449\\
NGC\,5490&-0.043&0.546&NGC\,3923&0.014&0.577\\
NGC\,6086&-0.06&0.511&NGC\,3998&0.062&1.844\\
NGC\,6251&0.148&2.054&NGC\,4026&-0.017&0.49\\
NGC\,7052&-0.01&1.536&NGC\,4036&0.013&1.093\\
NGC\,7626&-0.01&0.516&NGC\,4143&-0.024&1.101\\
UGC\,12064&-0.05&1.392&NGC\,4203&0.044&1.442\\
UGC\,1841&-0.011&0.952&NGC\,4281&0.009&1.719\\
UGC\,7115&-0.058&0.882&NGC\,4429&0.059&2.022\\
&&&NGC\,4459&0.043&2.085\\
&&&NGC\,4472&-0.366&0.601\\
&&&NGC\,4477&-0.021&0.891\\
&&&NGC\,4526&0.075&2.264\\
&&&NGC\,4546&-0.032&0.797\\
&&&NGC\,4552&-0.026&0.677\\
&&&NGC\,4564&-0.043&0.562\\
&&&NGC\,4596&-0.032&1.495\\
&&&NGC\,4636&-0.025&0.585\\
&&&NGC\,4649&-0.491&0.675\\
&&&NGC\,4697&-0.008&0.921\\
&&&NGC\,4889&-0.055&0.498\\
&&&NGC\,5018&0.05&1.294\\
&&&NGC\,5044&-0.025&0.83\\
&&&NGC\,5077&-0.044&0.779\\
&&&NGC\,5102&0.085&1.003\\
&&&NGC\,5252&0.829&2.018\\
&&&NGC\,5273&0.172&2.547\\
&&&NGC\,5328&-0.044&0.529\\
&&&NGC\,5813&0.004&0.414\\
&&&NGC\,5846&-0.031&0.558\\
&&&NGC\,6958&0.021&1.552\\
&&&NGC\,7332&-0.047&0.659\\
&&&NGC\,7619&-0.023&0.508\\
\hline
    \end{tabular}
    \label{tab:wise}
\end{table*}

\section{Far-infrared data} 
\label{sec:iras}

\begin{table*}[h!]
    \caption{Table containing information on measured far-infrared fluxes for all sources observed by the IRAS satellite. }
\tiny
    \centering
    \begin{tabular}{lcccccccc}
    \hline
    Name&$F_{60\mu m}$&$err\_F_{60\mu m}$&$F_{100\mu m}$&$err\_F_{100\mu m}$&$FIR$&$err\_FIR$&q&$err_q$\\
&[Jy]&[Jy]&[Jy]&[Jy]&[$10^{-14}$\,W\,m$^{-2}$]&[$10^{-14}$\,W\,m$^{-2}$]&-&-\\
(1)&(2)&(3)&(4)&(5)&(6)&(7)&(8)&(9)\\
\hline
3C\,120&1.32&0.09&2.64&0.18&7.56&0.38&-0.23&0.03\\
Ark\,120&0.66&0.07&1.30&0.14&3.74&0.28&1.91&0.04\\
MRK\,0279&1.08&0.17&2.28&0.18&6.33&0.61&1.85&0.05\\
MRK\,0509&1.42&0.11&1.43&0.20&6.37&0.45&1.96&0.04\\
NGC\,0383&0.44&0.05&1.20&0.20&2.93&0.31&-0.79&0.05\\
NGC\,0404&2.27&0.27&4.78&0.53&13.30&1.11&2.98&0.07\\
NGC\,0524&0.79&0.08&1.74&0.23&4.71&0.38&2.61&0.07\\
NGC\,0741$^a$&0.19&0.02&1.13&0.11&2.03&0.16&-0.29&0.03\\
NGC\,1052&0.94&0.08&1.22&0.13&4.55&0.3&0.12&0.03\\
NGC\,1275&7.22&0.65&8.01&0.72&33.3&2.3&-0.41&0.03\\
NGC\,1316$^a$&3.07&0.03&8.11&1.90&20.04&2.4&2.02&0.32\\
NGC\,1332&0.52&0.04&1.80&0.13&3.92&0.21&2.34&0.05\\
NGC\,1358$^a$&0.38&0.05&0.93&0.20&2.38&0.29&1.51&0.06\\
NGC\,1380&1.05&0.07&2.96&0.24&7.09&0.38&3.07&0.14\\
NGC\,1407$^a$&0.14&0.03&0.48&0.07&1.05&0.13&0.50&0.06\\
NGC\,1600$^a$&0.10&0.03&0.19&0.07&0.56&0.13&0.38&0.10\\
NGC\,2110&4.46&0.4&6.26&0.69&22.21&1.57&1.30&0.03\\
NGC\,2693&0.25&0.02&0.52&0.05&1.45&0.09&1.66&0.05\\
NGC\,2974&0.48&0.05&1.60&0.29&3.54&0.4&1.96&0.05\\
NGC\,3245&2.03&0.18&3.37&0.30&10.76&0.71&2.63&0.04\\
NGC\,3516&1.74&0.12&2.17&0.13&8.32&0.43&1.85&0.03\\
NGC\,3557$^a$&0.24&0.05&0.75&0.17&1.71&0.26&-0.24&0.07\\
NGC\,3665&2.00&0.24&6.37&0.64&14.41&1.12&1.53&0.04\\
NGC\,3706$^a$&0.07&0.05&0.22&0.08&0.50&0.18&0.40&0.16\\
NGC\,3801$^a$&0.17&0.05&2.80&0.09&4.05&0.20&-0.02&0.03\\
NGC\,3842$^a$&0.36&0.06&1.49&0.19&3.02&0.30&1.81&0.05\\
NGC\,3998&0.44&0.05&0.93&0.16&2.58&0.25&0.84&0.04\\
NGC\,4026$^a$&0.1&0.03&0.56&0.13&1.02&0.18&2.32&0.18\\
NGC\,4203&0.64&0.06&2.11&0.21&4.70&0.34&2.31&0.05\\
NGC\,4261$^a$&0.08&0.04&0.15&0.05&0.45&0.13&-2.27&0.13\\
NGC\,4278&0.57&0.07&1.62&0.16&3.85&0.30&0.43&0.04\\
NGC\,4281&0.59&0.07&1.55&0.53&3.85&0.70&3.11&0.28\\
NGC\,4429&1.54&0.18&4.62&0.55&10.74&0.92&3.41&0.05\\
NGC\,4459&1.75&0.21&4.34&0.52&11.07&0.95&3.33&0.04\\
NGC\,4477&0.54&0.06&1.18&0.19&3.21&0.32&3.03&0.27\\
NGC\,4526&5.93&0.71&16.00&2.24&39.12&3.65&2.94&0.04\\
NGC\,4546$^a$&0.26&0.05&0.89&0.22&1.95&0.31&1.67&0.07\\
NGC\,4552$^a$&0.16&0.05&0.53&0.06&1.18&0.17&0.50&0.06\\
NGC\,4596&0.49&0.09&1.28&0.18&3.17&0.36&3.02&0.28\\
NGC\,4649$^a$&0.78&0.03&1.09&0.07&3.88&0.14&1.55&0.02\\
NGC\,4697$^a$&0.46&0.02&1.24&0.08&3.03&0.12&3.13&0.36\\
NGC\,4786$^b$&0.28&0.05&0.73&0.17&1.83&0.26&1.72&0.07\\
NGC\,5018&1.01&0.11&1.90&0.19&5.63&0.43&2.88&0.11\\
NGC\,5044$^a$&0.14&0.06&0.15&0.07&0.64&0.20&0.67&0.14\\
NGC\,5102&0.83&0.07&2.65&0.21&5.99&0.36&2.97&0.13\\
NGC\,5128&172.00&15.48&338.00&27.04&977.20&60.77&0.80&0.05\\
NGC\,5273$^a$&0.90&0.04&1.56&0.12&4.85&0.19&2.57&0.05\\
NGC\,5283$^b$&0.13&0.03&0.75&0.14&1.36&0.20&1.43&0.07\\
NGC\,6958&1.06&0.10&2.45&0.25&6.48&0.44&2.03&0.04\\
NGC\,7052&0.45&0.04&1.48&0.12&3.29&0.21&0.73&0.03\\
NGC\,7332$^a$&0.21&0.03&0.41&0.11&1.19&0.17&2.72&0.37\\
\hline
    \end{tabular}
    \label{tab:iras}
\tablefoot{Column 1-- Object Name and references to the provided fluxes if not cataloged in IRAS Source catalogs: $^a$-- \citet{Knapp1989}; $^b$-- \citet{Moshir2008} \\Column 2-- $60 \mu m $ IRAS flux,\\ Column 4--$100 \mu m $ IRAS flux, \\Column 6-- Far-infrared flux calculated with the formula, $F_{\rm FIR}/{\rm W\,m^{-2}} = 1.26 \times 10^{-14} \, [2.58 \times S_{\rm 60\,\mu m}/{\rm Jy} + S_{\rm 100\,\mu m}/{\rm Jy}]$, \\Column 8-- fraction on FIR to 1.4\,GHz radio flux, \\Columns 3, 5, 7, and 9 contain the errors corresponding to Columns 2, 4, 6, and 8, respectively.}
\end{table*}

\begin{table*}[h!]
    \caption{Table containing information on the far-infrared fluxes for sources with only upper limit estimates as measured by the IRAS satellite.}
\tiny
    \centering
    \begin{tabular}{lclclclcl}
    \hline
    Name&$F_{60 \mu m}$&$err\_F_{60 \mu m}$&$F_{100 \mu m}$&$err\_F_{100\mu m}$&$FIR$&$err\_FIR$&q&$err_q$\\
    &[Jy]&[Jy]&[Jy]&[Jy]&[$10^{-14}$\,W\,m$^{-2}$]&[$10^{-14}$\,W\,m$^{-2}$]&-&-\\
(1)&(2)&(3)&(4)&(5)&(6)&(7)&(8)&(9)\\
\hline
PG\,0026+129$^d$&<0.03&-&<0.08&-&<0.19&-&<0.83&-\\
3C\,390.3&<0.4&-&<1&-&<2.54&-&< 1.22&-\\
Cygnus\,A&2.33&0.16&<8.28&-&<17.86&-&<0.47&-\\
IC\,1459&0.42&0.07&<1.04&-&<2.66&-&< -0.26&-\\
MRK\,1383&0.24&0.04&0.38&0.09&1.23&0.17&1.96&0.09\\
MRK\,335&0.45&0.05&<1&-&<2.71&-&<1.98&-\\
NGC\,0584$^c$&<0.04&-&0.59&0.1&0.87&0.13&<2.59&-\\
MRK\,771$^d$&0.16&0.04&<0.46&-&<1.1&-&<1.98&-\\
NGC\,0221&<0.17&-&<0.63&-&<1.35&-&<2.71&-\\
NGC\,0315&0.27&0.05&0.71&0.19&1.75&0.29&-0.22&0.07\\
NGC\,0541$^c$&<0.05&-&<0.13&-&<0.32&-&< -0.25&-\\
NGC\,1023&0.18&0.04&0.52&0.12&1.25&0.21&2.77&0.11\\
NGC\,1194&0.79&0.06&<1&-&<3.79&-&<2.61&-\\
NGC\,1399$^c$&<0.03&-&0.3&0.08&0.48&0.1&<-0.21&-\\
NGC\,1453$^c$&<0.04&-&0.75&0.13&1.06&0.16&<1&-\\
NGC\,1550$^c$&<0.03&-&<0.25&-&<0.39&-&<0.8&-\\
NGC\,2329$^c$&<0.03&-&<0.19&-&<0.35&-&< -0.6&-\\
NGC\,2787&0.66&0.07&<2.05&-&<4.68&-&<2.04&-\\
NGC\,2911&0.22&0.04&0.81&0.19&1.72&0.28&0.91&0.07\\
NGC\,3115$^c$&0.13&0.04&<0.11&-&<0.56&-&<2.4&-\\
NGC\,3379$^c$&<0.04&-&<0.11&-&<0.27&-&<1.42&-\\
NGC\,3414&0.19&0.04&0.91&0.21&1.76&0.29&2.03&0.08\\
NGC\,3585$^c$&0.16&0.04&<0.09&-&<0.63&-&<2.44&-\\
NGC\,3640$^c$&<0.04&-&<0.07&-&<0.21&-&<0.12&-\\
NGC\,3862$^c$&0.21&0.05&<0.18&-&<0.9&-&< -1.37&-\\
NGC\,3923$^c$&<0.04&-&<0.14&-&<0.28&-&<1.88&-\\
NGC\,4036&0.54&0.05&<1.6&-&<3.75&-&<1.94&-\\
NGC\,4151&<6.64&-&<8.73&-&<32.33&-&<1.38&-\\
NGC\,4335&0.27&0.04&1.09&0.24&2.22&0.33&0.69&0.07\\
NGC\,4374&0.51&0.05&<1.41&-&<3.39&-&< -0.89&-\\
NGC\,4472$^a$&<0.19&-&<0.48&-&<1.21&-&<0.17&-\\
NGC\,4486&0.39&0.05&1.02&0.3&2.55&0.41&-2.49&0.07\\
NGC\,4564$^c$&<0.06&-&<0.19&-&<0.43&-&<1.86&-\\
NGC\,4636$^c$&0.14&0.04&<0.07&-&<0.54&-&<0.26&-\\
NGC\,4889$^c$&<0.05&-&<0.07&-&<0.26&-&<1.76&-\\
NGC\,5127$^c$&<0.03&-&<0.06&-&<0.18&-&< -1.61&-\\
NGC\,5328$^c$&<0.03&-&<0.07&-&<0.19&-&<1.76&-\\
NGC\,5419$^c$&<0.04&-&0.23&0.07&0.41&0.08&<-0.51&-\\
NGC\,5490$^c$&<0.03&-&<0.19&-&<0.35&-&< -0.35&-\\
NGC\,5813$^c$&<0.02&-&<0.1&-&<0.2&-&< 0.53&-\\
NGC\,5846$^c$&<0.04&-&<0.13&-&<0.27&-&<0.54&-\\
NGC\,6086$^c$&<0.05&-&<0.19&-&<0.39&-&<0.01&-\\
NGC\,6251$^c$&<0.02&-&<0.11&-&<0.21&-&< -1.51&-\\
NGC\,7619$^c$&<0.04&-&<0.23&-&<0.41&-&<0.73&-\\
NGC\,7626$^c$&<0.04&-&<0.13&-&<0.29&-&< -1.03&-\\
UGC\,1841$^b$&<0.13&-&<0.35&-&<0.86&-&< -1.55&-\\
\hline
    \end{tabular}
\tablefoot{Column 1-- Object Name and references to the provided fluxes if not cataloged in the IRAS Source catalogs: $^a$--\citet{Rice1988}; $^b$-- \citet{Golombek1988}; $^c$-- \citet{Knapp1989}; $^d$-- \citet{Sanders1989},\\Column 2-- $60 \mu m $ IRAS flux,\\ Column 4--$100 \mu m $ IRAS flux, \\Column 6-- Far-infrared flux calculated with the formula, $F_{\rm FIR}/{\rm W\,m^{-2}} = 1.26 \times 10^{-14} \, [2.58 \times S_{\rm 60\,\mu m}/{\rm Jy} + S_{\rm 100\,\mu m}/{\rm Jy}]$, \\Column 8-- fraction on FIR to 1.4\,GHz radio flux, \\Columns 3, 5, 7, and 9 contain the errors corresponding to Columns 2, 4, 6, and 8, respectively.}
    \label{tab:irasupper}

\end{table*}

%\appendix

%\section{Radio and Infrared Data.} 
%\label{A:data}

%In Table\,\ref{tab:data}, we summarize the gathered $45^{\prime\prime}$-resolution data at 1.4\,GHz for the 62 radio-detected early-type galaxies from the list of \citet{gaspari19}. We note that, in addition to those 62 sources, nine early-type galaxies from \citeauthor{gaspari19} have only the VLA upper limits at the level of 0.5\,mJy \citep[NGC\,0821, NGC\,1023,  NGC\,1374, NGC\,3377, NGC\,4291, NGC\,4459, NGC\,4473, NGC\,4621, and NGC\,5576; see][]{brown2011}, while five are missing any arcmin-scale data at GHz frequencies (NGC\,1277, NGC\,4342, NGC\,5845, NGC\,6861, and NGC\,7768). The corresponding far-infrared IRAS fluxes, as well as mid-infrared WISE colours, are provided in the Table as well.

\clearpage
\section{The sample}
\label{sec:radiodata}

%==================== PART 1/4 ====================%
%\usepackage{pdflscape}

\begin{landscape}
%\onecolumn
\tiny
\setlength{\tabcolsep}{2.5pt} 
\begin{longtable}{lllllllllllll}

\caption{Radio and infrared data for radio-detected early-type galaxies \label{tab:radio}} \\
\hline
Name & RA & DEC & morph & $F_{\rm 1.4}$ & $F_{3}$ & Dist & $M_{BH}$ & Ref. & $log_{10}\sigma$ & Ref. & $log_{10}\frac{L_{1.4}}{L_{Edd}}$ & $log_{10}\frac{L_{3}}{L_{Edd}}$\\
 & [deg] & [deg] & & [mJy] & [mJy] & [Mpc] & [$M_\odot$] & & [km/s] & & &  \\
(1)&(2)&(3)&(4)&(5)&(6)&(7)&(8)&(9)&(10)&(11)&(12)&(13)\\
\hline
\endfirsthead
\multicolumn{13}{c}%
{{\tablename\ \thetable{} -- continued from previous page}} \\
\hline
Name & RA & DEC & morph & $F_{\rm 1.4}$ & $F_{3}$ & Dist & $M_{BH}$ & Ref. & $\log_{10}\sigma$ & Ref. & $log_{10}\frac{L_{1.4}}{L_{Edd}}$ & $log_{10}\frac{L_{3}}{L_{Edd}}$\\
 & [deg] & [deg] & & [mJy] & [mJy] & [Mpc] & [$M_\odot$] & & [km/s] & & &  \\
(1)&(2)&(3)&(4)&(5)&(6)&(7)&(8)&(9)&(10)&(11)&(12)&(13)\\
\hline
\endhead
\hline \multicolumn{13}{r}{{Continued on next page}} \\
\endfoot
\hline
\endlastfoot
0026+129&7.307083&13.267778&E&7.4$\pm$0.5$^b$&1.2&406.2$\pm$40.62&8.49$^{+0.10}_{-0.12}$&[B18]&-&&-7.28&-7.74\\
3C\,120&68.29596&5.35428&L&3439$\pm$103.2$^b$&2615.3&141.4$\pm$14.1&7.73$^{+0.15}_{-0.15}$&[B16]&2.21$\pm$0.05&[B16]&-4.77&-4.56\\
3C\,390.3&280.537458&79.771424&E&11226$\pm$1122.6$^a$&87.4&240.3$\pm$24&8.62$^{+0.16}_{-0.16}$&[B16]&2.44$\pm$0.03&[B16]&-4.68&-6.46\\
A\,1836\,BCG&210.424356&-11.607026&L&1870$\pm$39$^f$&24.2&152.4$\pm$8.4&9.57$^{+0.06}_{-0.06}$&[B16]&2.46$\pm$0.02&[B16]&-6.81&-8.37\\
Ark120&79.04763&-0.15006&L&12.3$\pm$0.6$^b$&1.8&140.1$\pm$14&8.05$^{+0.17}_{-0.17}$&[B16]&2.28$\pm$0.02&[B16]&-7.54&-8.05\\
Arp151&71.400697&54.382502&L&11.2$\pm$0.5$^b$&0.4&90.3$\pm$9&6.65$^{+0.16}_{-0.16}$&[B16]&2.07$\pm$0.01&[B16]&-6.57&-7.68\\
Cygnus\,A&299.868153&40.733916&L&1598$\pm$41$^d$&-&257.1$\pm$25.7&9.41$^{+0.13}_{-0.13}$&[B16]&2.43$\pm$0.02&[B16]&-6.26&-\\
ESO 444- G 025&201.027727&-31.670042&L&8.8$\pm$0.6$^b$&1.1&149.4$\pm$14.94&9.98$^{+0.02}_{-0.02}$&[B21]&-&&-9.56&-10.14\\
Holm 15A&10.460294&-9.303129&E&58.1$\pm$2.5$^b$&6.6&163.9$\pm$16.39&10.60$^{+0.09}_{-0.09}$&[M19]&2.45$\pm$0.05&[V18]&-9.29&-9.90\\
IC1459&344.29446&-36.46194&E&1279.7$\pm$45.2$^b$&317.5&28.9$\pm$3.7&9.39$^{+0.08}_{-0.08}$&[B16]&2.53$\pm$0.02&[B16]&-8.24&-8.51\\
Mrk0279&208.264362&69.308213&L&24$\pm$1.1$^b$&3.9&130.4$\pm$13&7.40$^{+0.23}_{-0.23}$&[B16]&2.29$\pm$0.03&[B16]&-6.67&-7.12\\
Mrk0509&311.04033&-10.72306&L&18.6$\pm$1$^b$&2.9&147.3$\pm$14.7&8.03$^{+0.15}_{-0.15}$&[B16]&2.26$\pm$0.03&[B16]&-7.30&-7.78\\
Mrk1216&127.196292&-6.940139&E&9.2$\pm$0.6$^b$&2.5&94$\pm$9.4&9.69$^{+0.15}_{-0.15}$&[S19]&2.51$\pm$0.05&[S19]&-9.66&-9.89\\
Mrk1310&180.30982&-3.67808&E&3.4$\pm$0.5$^b$&0.7&83.8$\pm$8.4&6.19$^{+0.19}_{-0.19}$&[B16]&1.92$\pm$0.03&[B16]&-6.69&-7.04\\
Mrk1383&217.27745&1.285132&E&3.6$\pm$0.6$^b$&0.4&254.4$\pm$25.44&8.65$^{+0.13}_{-0.13}$&[Y19]&2.34$\pm$0.03&[W08]&-8.16&-8.78\\
Mrk0335&1.581339&20.202914&E&7.6$\pm$0.6$^b$&1.9&76.9$\pm$7.69&7.23$^{+0.04}_{-0.04}$&[B18]&-&&-7.45&-7.72\\
Mrk0771&188.01502&20.158114&L&3.1$\pm$0.5$^b$&1.8&187.8$\pm$18.78&8.09$^{+0.37}_{-0.37}$&[Y19]&2.21$\pm$0.09&[D07]&-7.93&-7.83\\
NGC0193&9.827446&3.331117&E&983.1$\pm$31$^b$&19.7&49.7$\pm$5&8.40$^{+0.32}_{-0.32}$&[B16]&2.27$\pm$0.04&[B16]&-6.89&-8.26\\
NGC0221&10.6743&40.865287&E&0.7$\pm$0.5$^e$&-&0.8$\pm$0.01&6.39$^{+0.19}_{-0.19}$&[B16]&1.87$\pm$0.02&[B16]&-11.61&-\\
NGC0315&14.45368&30.352448&E&1800$\pm$100$^e$&213.3&57.7$\pm$5.8&8.92$^{+0.31}_{-0.31}$&[B16]&2.49$\pm$0.04&[B16]&-7.02&-7.61\\
NGC0383&16.853995&32.412559&L&4800$\pm$200$^e$&33.4&59.2$\pm$5.9&8.76$^{+0.32}_{-0.32}$&[B16]&2.38$\pm$0.03&[B16]&-6.41&-8.24\\
NGC0404&17.362587&35.718131&E&3.7$\pm$0.5$^b$&0.6&3.1$\pm$0.3&5.65$^{+0.25}_{-0.25}$&[B16]&1.39$\pm$0.09&[B16]&-8.98&-9.43\\
NGC0524&21.19867&9.53936&L&3.1$\pm$0.4$^b$&0.6&24.2$\pm$2.2&8.94$^{+0.05}_{-0.05}$&[B16]&2.37$\pm$0.02&[B16]&-10.56&-10.94\\
NGC0541&21.434617&-1.37958&E&538.2$\pm$11.9$^f$&3.6&63.7$\pm$6.4&8.59$^{+0.34}_{-0.34}$&[B16]&2.28$\pm$0.01&[B16]&-7.13&-8.97\\
NGC0584&22.836458&-6.868056&E&0.6$\pm$0.5$^e$&-&18.3$\pm$1.83&8.13$^{+0.17}_{-0.17}$&[T19]&2.28$\pm$0.01&[T19]&-10.70&-\\
NGC0741&29.087639&5.628938&E&1059$\pm$21$^f$&4.9&65.7$\pm$6.6&8.67$^{+0.37}_{-0.37}$&[B16]&2.37$\pm$0.02&[B16]&-6.89&-8.89\\
NGC1023&40.100042&39.063285&L&0.6$\pm$0.1$^g$&-&10.8$\pm$0.8&7.62$^{+0.05}_{-0.05}$&[B16]&2.22$\pm$0.02&[B16]&-10.68&-\\
NGC1052&40.26996&-8.25586&E&912.5$\pm$27.4$^b$&376.4&18.1$\pm$1.8&8.24$^{+0.29}_{-0.29}$&[B16]&2.28$\pm$0.01&[B16]&-7.64&-7.69\\
NGC1194&45.954621&-1.103743&L&2.5$\pm$0.2$^c$&0.6&58$\pm$6.3&7.85$^{+0.05}_{-0.05}$&[B16]&2.17$\pm$0.07&[B16]&-8.80&-9.09\\
NGC1275&49.950667&41.511696&E&22829.2$\pm$684.9$^b$&9319.1&70$\pm$7&8.98$^{+0.20}_{-0.20}$&[B16]&2.39$\pm$0.05&[B16]&-5.81&-5.87\\
NGC1316&27.823171&34.848618&E&150000$\pm$10000$^b$&19.6&18.6$\pm$0.6&8.18$^{+0.25}_{-0.25}$&[B16]&2.35$\pm$0.02&[B16]&-5.34&-8.89\\
NGC1332&51.571884&-21.335216&E&4.8$\pm$0.5$^b$&0.9&22.3$\pm$1.9&8.83$^{+0.04}_{-0.04}$&[B16]&2.52$\pm$0.02&[B16]&-10.33&-10.72\\
NGC1358&53.41562&-5.08956&L&19.6$\pm$0.7$^b$&1.1&48.2$\pm$4.8&8.37$^{+0.32}_{-0.32}$&[B16]&2.23$\pm$0.05&[B16]&-8.59&-9.51\\
NGC1380&54.114968&-34.976225&L&1.6$\pm$0.5$^e$&0.7&18.7$\pm$1.87&8.17$^{+0.17}_{-0.17}$&[K22]&2.34$\pm$0.02&[V11]&-10.29&-10.33\\
NGC1399&54.62092&-35.45019&E&2200$\pm$100$^e$&5.4&20.9$\pm$0.7&8.94$^{+0.31}_{-0.31}$&[B16]&2.53$\pm$0.02&[B16]&-7.83&-10.11\\
NGC1407&55.04971&-18.58028&E&87.7$\pm$3.5$^b$&5.7&28$\pm$3.4&9.65$^{+0.08}_{-0.08}$&[B16]&2.45$\pm$0.02&[B16]&-9.69&-10.55\\
NGC1453&56.613542&-3.968778&E&28.1$\pm$1$^b$&8&38.7$\pm$3.87&9.46$^{+0.06}_{-0.06}$&[L20]&2.43$\pm$0.05&[B16]&-9.71&-9.93\\
NGC1497&60.52846&23.13292&L&22.8$\pm$0.8$^b$&12.9&75.3$\pm$7.5&8.63$^{+0.19}_{-0.19}$&[B16]&2.39$\pm$0.04&[B16]&-8.39&-8.31\\
NGC1550&64.90817&2.40992&E&16.6$\pm$1.6$^b$&0.2&51.6$\pm$5.6&9.57$^{+0.07}_{-0.07}$&[B16]&2.48$\pm$0.02&[B16]&-9.80&-11.39\\
NGC1600&67.91583&-5.08678&E&61.6$\pm$2.6$^b$&0.5&64$\pm$6.4&10.23$^{+0.04}_{-0.04}$&[B16]&2.47$\pm$0.02&[B16]&-9.70&-11.46\\
NGC2110&88.04725&-7.45628&E&298.8$\pm$9$^b$&60.8&29.1$\pm$2.9&8.12$^{+0.64}_{-0.64}$&[B16]&2.30$\pm$0.05&[B16]&-7.59&-7.95\\
NGC2179&92.00933&-21.74714&L&13.8$\pm$0.7$^b$&2.1&35.8$\pm$3.6&8.31$^{+0.23}_{-0.23}$&[B16]&2.19$\pm$0.03&[B16]&-8.94&-9.43\\
NGC2329&107.28262&48.61478&E&363.7$\pm$13.1$^b$&52.5&72.3$\pm$7.2&8.18$^{+0.18}_{-0.18}$&[B16]&2.34$\pm$0.03&[B16]&-6.78&-7.29\\
NGC2685&133.894628&58.734398&L&1.9$\pm$0.5$^e$&-&12.5$\pm$1.3&6.59$^{+0.41}_{-6.59}$&[B16]&2.02$\pm$0.02&[B16]&-8.99&-\\
NGC2693&134.246955&51.34744&E&8.5$\pm$0.9$^b$&1.2&48$\pm$4.8&9.23$^{+0.10}_{-0.10}$&[P22]&2.54$\pm$0.06&[K01]&-9.81&-10.33\\
NGC2787&139.827486&69.203253&L&11.3$\pm$0.6$^b$&4.7&7.4$\pm$1.2&7.61$^{+0.09}_{-0.09}$&[B16]&2.28$\pm$0.02&[B16]&-9.69&-9.74\\
NGC2892&143.220542&67.617389&E&236$\pm$23.6$^c$&8.8&86.2$\pm$8.6&8.43$^{+0.11}_{-0.11}$&[B16]&2.47$\pm$0.03&[B16]&-7.06&-8.16\\
NGC2911&143.44212&10.15261&E&56.5$\pm$1.7$^b$&23.1&43.5$\pm$4.3&9.09$^{+0.29}_{-0.29}$&[B16]&2.32$\pm$0.03&[B16]&-8.94&-8.99\\
NGC2974&145.63825&-3.69969&L&10.4$\pm$0.5$^b$&2.6&21.5$\pm$2.4&8.23$^{+0.09}_{-0.09}$&[B16]&2.36$\pm$0.02&[B16]&-9.42&-9.70\\
NGC3078&149.60254&-26.92672&E&313.1$\pm$10.9$^b$&47.9&32.8$\pm$3.3&7.91$^{+0.42}_{-0.42}$&[B16]&2.32$\pm$0.03&[B16]&-7.26&-7.74\\
NGC3091&150.059543&-19.636961&E&2.5$\pm$0.5$^e$&0.4&51.2$\pm$8.3&9.56$^{+0.07}_{-0.07}$&[B16]&2.49$\pm$0.02&[B16]&-10.62&-11.08\\
NGC3115&151.30825&-7.718583&L&0.6$\pm$0.5$^e$&-&9.5$\pm$0.4&8.95$^{+0.09}_{-0.09}$&[B16]&2.36$\pm$0.02&[B16]&-12.09&-\\
NGC3245&156.82575&28.50756&E&6.7$\pm$0.5$^e$&1.6&21.4$\pm$2&8.38$^{+0.11}_{-0.11}$&[B16]&2.25$\pm$0.02&[B16]&-9.77&-10.06\\
NGC3379&161.956616&12.581624&E&2.7$\pm$0.5$^b$&0.3&10.7$\pm$0.5&8.62$^{+0.11}_{-0.11}$&[B16]&2.27$\pm$0.02&[B16]&-11.01&-11.63\\
NGC3414&162.81825&27.97492&L&4.4$\pm$0.4$^e$&1.6&25.2$\pm$2.7&8.40$^{+0.07}_{-0.07}$&[B16]&2.28$\pm$0.02&[B16]&-9.83&-9.94\\
NGC3489&165.077375&13.901222&L&1.5$\pm$0.5$^e$&-&12.1$\pm$0.8&6.78$^{+0.05}_{-0.05}$&[B16]&2.01$\pm$0.02&[B16]&-9.31&-\\
NGC3516&166.69879&72.56953&L&31.3$\pm$1.3$^b$&2.1&37.9$\pm$3.8&7.37$^{+0.16}_{-0.16}$&[B16]&2.26$\pm$0.01&[B16]&-7.59&-8.44\\
NGC3557&167.490175&-37.539155&E&790$\pm$60$^e$&9.7&30.7$\pm$3.07&8.85$^{+0.02}_{-0.02}$&[R19]&2.45$\pm$0.03&[B07]&-7.85&-9.43\\
NGC3585&168.321214&-26.75484&E&0.6$\pm$0.5$^e$&-&20.5$\pm$1.7&8.52$^{+0.13}_{-0.13}$&[B16]&2.33$\pm$0.02&[B16]&-10.99&-\\
NGC3607&169.22775&18.05125&E&6.9$\pm$0.4$^b$&1.3&22.6$\pm$1.8&8.14$^{+0.16}_{-0.16}$&[B16]&2.32$\pm$0.02&[B16]&-9.47&-9.86\\
NGC3608&169.245632&18.148684&E&1.3$\pm$0.5$^e$&-&22.8$\pm$1.5&8.67$^{+0.10}_{-0.10}$&[B16]&2.23$\pm$0.02&[B16]&-10.72&-\\
NGC3640&170.278542&3.234833&L&42$\pm$2$^e$&-&13$\pm$1.3&8.89$^{+0.02}_{-0.02}$&[T19]&2.25$\pm$0.02&[T19]&-9.92&-\\
NGC3665&171.18212&38.76281&E&112.2$\pm$3.7$^b$&3.8&34.7$\pm$6.7&8.76$^{+0.09}_{-0.09}$&[B16]&2.34$\pm$0.02&[B16]&-8.51&-9.64\\
NGC3706&172.435125&-36.391306&L&53$\pm$2$^e$&0.9&46$\pm$4.6&9.77$^{+0.06}_{-0.06}$&[B16]&2.51$\pm$0.01&[B16]&-9.60&-11.04\\
NGC3801&175.07096&17.7275&L&1143.3$\pm$38.5$^f$&7&46.3$\pm$4.6&8.28$^{+0.31}_{-0.31}$&[B16]&2.32$\pm$0.04&[B16]&-6.77&-8.65\\
NGC3842&176.008958&19.949806&E&12.6$\pm$0.6$^b$&0.5&92.2$\pm$10.6&9.96$^{+0.14}_{-0.14}$&[B16]&2.44$\pm$0.02&[B16]&-9.81&-10.88\\
NGC3862&176.270871&19.606317&E&5689$\pm$131$^f$&139&84.6$\pm$8.5&8.41$^{+0.37}_{-0.37}$&[B16]&2.32$\pm$0.03&[B16]&-5.68&-6.96\\
NGC3923&177.757059&-28.806017&E&1$\pm$0.5$^e$&-&20.9$\pm$2.7&9.45$^{+0.12}_{-0.12}$&[B16]&2.35$\pm$0.02&[B16]&-11.69&-\\
NGC3945&178.307208&60.675556&L&1.6$\pm$0.5$^e$&1.1&19.5$\pm$2&6.94$^{+0.46}_{-6.94}$&[B16]&2.25$\pm$0.02&[B16]&-9.03&-8.86\\
NGC3998&179.48392&55.4535&E&98.4$\pm$3$^b$&33.3&14.3$\pm$1.3&8.93$^{+0.05}_{-0.05}$&[B16]&2.35$\pm$0.02&[B16]&-9.50&-9.64\\
NGC4026&179.854958&50.961694&L&1.3$\pm$0.5$^e$&-&13.4$\pm$1.7&8.26$^{+0.12}_{-0.12}$&[B16]&2.19$\pm$0.02&[B16]&-10.77&-\\
NGC4036&180.35925&61.89569&L&11.6$\pm$0.5$^b$&2.2&19$\pm$1.9&7.89$^{+0.36}_{-0.36}$&[B16]&2.26$\pm$0.02&[B16]&-9.14&-9.54\\
NGC4143&182.40158&42.5335&L&9.9$\pm$0.9$^b$&1.4&14.8$\pm$1.5&7.92$^{+0.36}_{-0.36}$&[B16]&2.25$\pm$0.02&[B16]&-9.46&-9.98\\
NGC4150&182.640247&30.401615&E&1.2$\pm$0.5$^e$&-&12.8$\pm$1.3&5.94$^{+0.44}_{-5.94}$&[B16]&1.91$\pm$0.02&[B16]&-8.52&-\\
NGC4151&182.6355&39.40581&L&359.6$\pm$10.8&70.9&20$\pm$2.8&7.81$^{+0.08}_{-0.08}$&[B16]&1.98$\pm$0.01&[B16]&-7.53&-7.90\\
NGC4203&183.77179&33.19797&L&6.1$\pm$0.5$^b$&2.7&14.1$\pm$1.4&7.82$^{+0.26}_{-0.26}$&[B16]&2.11$\pm$0.02&[B16]&-9.61&-9.64\\
NGC4261&184.846752&5.825215&E&22000$\pm$1000$^e$&111.4&32.4$\pm$2.8&8.72$^{+0.10}_{-0.10}$&[B16]&2.42$\pm$0.02&[B16]&-6.23&-8.20\\
NGC4278&185.02821&29.28081&E&385$\pm$11.6$^b$&105.5&15$\pm$1.5&7.96$^{+0.27}_{-0.27}$&[B16]&2.33$\pm$0.02&[B16]&-7.90&-8.13\\
NGC4281&185.089684&5.386392&L&0.8$\pm$0.5$^e$&-&26.7$\pm$1.5&8.73$^{+0.08}_{-0.08}$&[T19]&2.36$\pm$0.02&[T19]&-10.85&-\\
NGC4335&185.7585&58.44464&E&121.3$\pm$4.5$^b$&8.6&59.1$\pm$5.9&8.39$^{+0.31}_{-0.31}$&[B16]&2.41$\pm$0.01&[B16]&-7.64&-8.46\\
NGC4374&186.265597&12.886983&E&7000$\pm$600$^e$&52.8&10.2$\pm$1.02&8.97$^{+0.05}_{-0.05}$&[S16]&2.47$\pm$0.02&[S16]&-7.98&-9.78\\
NGC4382&186.350451&18.191487&E&0.7$\pm$0.5$^e$&-&17.9$\pm$1.8&7.11$^{+1.24}_{-7.11}$&[B16]&2.25$\pm$0.02&[B16]&-9.64&-\\
NGC4429&186.86045&11.10771&L&1.1$\pm$0.1$^g$&0.3&18.2$\pm$1.8&7.85$^{+0.35}_{-0.35}$&[B16]&2.25$\pm$0.02&[B16]&-10.16&-10.40\\
NGC4459&187.250037&13.978373&E&1.4$\pm$0.1$^g$&0.4&16$\pm$0.5&7.84$^{+0.09}_{-0.09}$&[B16]&2.20$\pm$0.02&[B16]&-10.16&-10.37\\
NGC4472&187.44417&8.00081&E&219.9$\pm$7.8$^b$&22.5&17.1$\pm$0.6&9.4$^{+0.04}_{-0.04}$&[B16]&2.40$\pm$0.02&[B16]&-9.47&-10.13\\
NGC4477&187.509159&13.636604&L&0.8$\pm$0.5$^e$&-&20.8$\pm$2.1&7.55$^{+0.30}_{-0.30}$&[B16]&2.17$\pm$0.02&[B16]&-9.89&-\\
NGC4486&187.70593&12.391123&E&210000$\pm$10000$^e$&1062.3&16.7$\pm$0.6&9.58$^{+0.10}_{-0.10}$&[B16]&2.42$\pm$0.02&[B16]&-6.69&-8.65\\
NGC4526&188.51158&7.69944&L&12$\pm$0.5$^b$&-&16.4$\pm$1.8&8.65$^{+0.12}_{-0.12}$&[B16]&2.32$\pm$0.02&[B16]&-10.02&-\\
NGC4546&188.872958&-3.793194&E&11$\pm$0.6$^b$&4.2&10.6$\pm$1.06&8.41$^{+0.03}_{-0.03}$&[R20]&2.40$\pm$0.02&[R16]&-10.19&-10.28\\
NGC4552&188.91575&12.55617&E&100.1$\pm$3$^b$&26&15.3$\pm$1&8.70$^{+0.05}_{-0.05}$&[B16]&2.35$\pm$0.02&[B16]&-9.21&-9.46\\
NGC4564&189.112428&11.439283&E&1.6$\pm$0.5$^e$&-&15.9$\pm$0.5&7.95$^{+0.12}_{-0.12}$&[B16]&2.19$\pm$0.02&[B16]&-10.22&-\\
NGC4596&189.98311&10.17614&L&0.8$\pm$0.5$^e$&-&16.5$\pm$6.2&7.89$^{+0.26}_{-0.26}$&[B16]&2.10$\pm$0.02&[B16]&-10.43&-\\
NGC4636&190.707615&2.687776&E&78.7$\pm$2.9$^b$&3.4&13.7$\pm$1.4&8.58$^{+0.22}_{-0.22}$&[B16]&2.26$\pm$0.02&[B16]&-9.29&-10.32\\
NGC4649&190.91721&11.55261&E&29.1$\pm$1.3$^b$&6.6&16.5$\pm$0.6&9.67$^{+0.10}_{-0.10}$&[B16]&2.43$\pm$0.02&[B16]&-10.65&-10.96\\
NGC4697&192.149491&-5.800742&L&0.6$\pm$0.5$^e$&1.4&12.5$\pm$0.4&8.31$^{+0.11}_{-0.11}$&[B16]&2.23$\pm$0.02&[B16]&-11.21&-10.52\\
NGC4786&193.635083&-6.859417&E&9.2$\pm$0.5$^e$&0.9&55.7$\pm$5.57&8.70$^{+0.33}_{-0.31}$&[K24]&2.45$\pm$0.04&[B16]&-9.12&-9.80\\
NGC4889&195.033875&27.977&E&1.2$\pm$0.5$^e$&0.3&102$\pm$5.2&10.32$^{+0.44}_{-0.44}$&[B16]&2.56$\pm$0.02&[B16]&-11.10&-11.37\\
NGC5018&198.254305&-19.518193&L&2$\pm$0.5$^e$&0.3&40.5$\pm$4.9&8.02$^{+0.08}_{-0.08}$&[B16]&2.32$\pm$0.01&[B16]&-9.38&-9.87\\
NGC5044&198.849875&-16.385528&E&36.1$\pm$1.5$^b$&8.4&27.8$\pm$2.78&9.28$^{+0.01}_{-0.01}$&[D17]&2.48$\pm$0.02&[D17]&-9.71&-10.01\\
NGC5077&199.88196&-12.65692&E&156.7$\pm$4.7$^b$&107&38.7$\pm$8.4&8.93$^{+0.27}_{-0.27}$&[B16]&2.35$\pm$0.02&[B16]&-8.44&-8.27\\
NGC5102&200.490029&-36.630244&L&1.7$\pm$0.5$^e$&-&4.7$\pm$0.47&5.94$^{+0.38}_{-0.38}$&[S19]&1.79$\pm$0.05&[S19]&-9.25&-\\
NGC5127&200.937583&31.56575&E&1980$\pm$198$^c$&4&62.5$\pm$6.3&8.27$^{+0.41}_{-0.41}$&[B16]&2.29$\pm$0.11&[B16]&-6.26&-8.62\\
NGC5128&201.365063&-43.019113&L&1330000$\pm$13300$^0$&-&3.6$\pm$0.2&7.76$^{+0.08}_{-0.08}$&[B16]&2.18$\pm$0.02&[B16]&-5.40&-\\
NGC5252&204.566&4.54364&L&16.3$\pm$0.6$^b$&3.9&103.7$\pm$10.4&9.07$^{+0.34}_{-0.34}$&[B16]&2.28$\pm$0.02&[B16]&-8.70&-8.99\\
NGC5273&205.53474&35.654214&E&3.5$\pm$0.4$^e$&0.6&15.5$\pm$1.5&6.61$^{+0.27}_{-0.27}$&[B16]&1.82$\pm$0.02&[B16]&-8.56&-9.00\\
NGC5283&205.273995&67.672311&L&13.4$\pm$0.6$^b$&2.6&34.5$\pm$3.5&7.41$^{+0.33}_{-0.33}$&[B16]&2.13$\pm$0.04&[B16]&-8.08&-8.46\\
NGC5328&208.222138&-28.489402&E&0.9$\pm$0.5$^e$&-&64.1$\pm$7&9.67$^{+0.16}_{-0.16}$&[B16]&2.52$\pm$0.02&[B16]&-10.98&-\\
NGC5419&210.91129&-33.97969&E&790$\pm$60$^e$&21.7&56.2$\pm$6.1&9.86$^{+0.14}_{-0.14}$&[B16]&2.57$\pm$0.01&[B16]&-8.34&-9.57\\
NGC5490&212.48908&17.54558&E&1300$\pm$100$^e$&21.8&65.2$\pm$6.5&8.73$^{+0.35}_{-0.35}$&[B16]&2.41$\pm$0.04&[B16]&-6.86&-8.31\\
NGC5813&225.296796&1.701981&E&15.8$\pm$1.1$^b$&1.2&19.7$\pm$1.97&8.85$^{+0.06}_{-0.06}$&[S16]&2.36$\pm$0.02&[S16]&-9.94&-10.73\\
NGC5846&226.62217&1.60528&E&21$\pm$1.3$^b$&48.7&24.9$\pm$2.3&9.04$^{+0.06}_{-0.06}$&[B16]&2.35$\pm$0.02&[B16]&-9.80&-9.11\\
NGC6086&243.148053&29.484791&E&100$\pm$1.4$^e$&-&138$\pm$11.5&9.57$^{+0.17}_{-0.17}$&[B16]&2.50$\pm$0.02&[B16]&-8.17&-\\
NGC6251&248.133208&82.537889&E&1800$\pm$100$^e$&174.8&108.4$\pm$9&8.79$^{+0.16}_{-0.16}$&[B16]&2.46$\pm$0.02&[B16]&-6.34&-7.02\\
NGC6958&312.177458&-37.997417&L&16.3$\pm$1$^b$&2.1&27.1$\pm$2.71&8.56$^{+0.33}_{-0.29}$&[T22]&2.23$\pm$0.01&[T22]&-9.36&-9.92\\
NGC7052&319.637708&26.447028&E&162.1$\pm$5.3$^b$&17.6&70.4$\pm$8.4&8.60$^{+0.23}_{-0.23}$&[B16]&2.42$\pm$0.02&[B16]&-7.57&-8.20\\
NGC7332&339.35225&23.798333&L&0.6$\pm$0.5$^e$&-&21.7$\pm$2.2&7.08$^{+0.18}_{-0.18}$&[B16]&2.10$\pm$0.02&[B16]&-9.51&-\\
NGC7619&350.06092&8.20633&E&20.3$\pm$1.1$^b$&4.7&51.5$\pm$7.4&9.40$^{+0.11}_{-0.11}$&[B18]&2.51$\pm$0.02&[B16]&-9.55&-9.85\\
NGC7626&350.177275&8.216967&E&842$\pm$15$^f$&10.4&38.1$\pm$3.8&8.58$^{+0.33}_{-0.33}$&[B16]&2.37$\pm$0.02&[B16]&-7.37&-8.95\\
UGC12064&337.83575&39.358194&L&3800$\pm$380$^c$&9.7&34.7$\pm$6.7&8.84$^{+0.52}_{-0.52}$&[B16]&2.41$\pm$0.03&[B16]&-7.06&-9.32\\
UGC1214&25.99042&2.34975&L&24$\pm$1.1$^b$&2.9&59.9$\pm$6&7.74$^{+0.18}_{-7.74}$&[B16]&2.02$\pm$0.06&[B16]&-7.68&-8.27\\
UGC1841&35.797547&42.992051&E&8200$\pm$820$^c$&82.9&74.9$\pm$7.5&8.47$^{+0.19}_{-0.19}$&[B16]&2.47$\pm$0.04&[B16]&-5.68&-7.35\\
UGC7115&182.0245&25.23672&E&87.3$\pm$2.7$^b$&17.7&88.2$\pm$8.8&9.00$^{+0.45}_{-0.45}$&[B16]&2.25$\pm$0.08&[B16]&-8.04&-8.41\\
UGC9799&229.18575&7.02169&E&5499.3$\pm$209.1$^b$&167.8&151.1$\pm$15.1&8.89$^{+0.8}_{-8.89}$&[B16]&2.37$\pm$0.02&[B16]&-5.67&-6.85\\
\end{longtable}
\tablefoot{\\
Column 1-- Source Name; \\
Column 2- and Column 3- RA and DEC J2000, in units of degrees;\\
Column 4-- host galaxy optical morphology: E-- elliptical, L-- lenticular;\\
Column 5-- 1.4 GHz radio flux, in mJy, References:$^0$~\citet{Cooper1965}; $^a$~\citet{White1992};
$^b$~\citet{Condon1998}; $^c$~\citet{Condon2002}; $^d$~\citet{Birzan2004};
$^e$~\citet{Brown2011}; $^f$~\citet{Allison2014}; $^g$~\citet{Nyland2017};\\
Column 6-- 3\,GHz core radio flux, in mJy, measured at the position of the optical core;\\
Column 7-- reported distance in Mpc;\\
Column 8-- logarithm of Black Hole Mass in $M_{\odot}$\\
Column 9-- references to Column 8: [B16]-- \citet{Bosch2016}; [S16]--;\citet{Saglia2016};[D17]--\citet{Diniz2017};[B18]--\citet{Bentz18}; [M19]--\citet{Mehrgan2019};[S19]--\citet{Sahu2019b}; [T19]--\citet{Thater2019};[Y19]--\citet{Yu2019};[L20]--\citet{Liepold2020};[R20]--\citet{Ricci2020};[B21]--\citet{denBrok2021};[K22]--\citet{Kabasares2022} ; [P22]--\citet{Pilawa2022};[T22]--\citet{Thater2022}\\
Column 10-- stellar Velocity dispersion, in km/s, measured in host galaxy effective radius $R_{eff}$\\
Column 11-- references to Column 10: [K01]-- \citet{Kuntschner2001};[B07]--\citet{Brough2007};[D07]-- \citet{Dasyra2007};[W08]--\citet{Watson2008}; [V11]--\citet{Vanderbeke2011};[B16]--\citet{Bosch2016} ;[R16]--\citet{Ricci2016};[S16]--\citet{Saglia2016};[D17]--\citet{Diniz2017};[V18]--\citet{Veale2018};[T19]--\citet{Thater2019}; [S19]--\citet{Sahu2019b};[T22]--\citet{Thater2022};\\
Column 12-- fraction of total 1.4\,GHz radio luminosity to Eddington luminosity ;\\
Column 13-- fraction of measured core-related 3\,GHz radio luminosity to Eddington luminosity ;
}

\end{landscape}

\end{document}